\UseRawInputEncoding
\pdfoutput=1

\documentclass[showpacs,nofootinbib,showkeys,eqsecnum,prd,aps,preprint]{revtex4-2}

\usepackage[english]{babel}
\usepackage{comment}

\usepackage{amssymb}
\usepackage{amsmath}
\usepackage{amsfonts}
\usepackage{latexsym}
\usepackage{mathrsfs}
\usepackage{bm}
\usepackage{tensor}
\usepackage{hyperref}
\usepackage{graphicx}

\def\const{\mathop{\rm const}\nolimits\,}
\def\dac{\displaystyle\frac}
\def\dil{\displaystyle\int\limits}

\usepackage{amsthm}
\usepackage{xcolor}

\newtheorem{defin}{Definition}[section]

\usepackage{diagbox}

\def\pa{\partial}

\def\BFC{{\bf C}}

\def\const{\mathop{\rm const}\nolimits\,}

\def\dac{\displaystyle\frac}

\def\dil{\displaystyle\int\limits}
\def\{{\lbrace}
\def\}{\rbrace}

\def\Or{{\rm O}}

\begin{document}

\title{Quasiparticle solutions for the nonlocal NLSE with an anti-Hermitian term in semiclassical approximation}

\author{Anton E. Kulagin}
\email{aek8@tpu.ru}
\affiliation{Tomsk Polytechnic University, 30 Lenina av., 634050 Tomsk, Russia}
\affiliation{V.E. Zuev Institute of Atmospheric Optics, SB RAS, 1 Academician Zuev Sq., 634055 Tomsk, Russia}

\author{Alexander V. Shapovalov}
\email{shpv@mail.tsu.ru}
\affiliation{Department of Theoretical Physics, Tomsk State University, Novosobornaya Sq. 1, 634050 Tomsk, Russia}
\affiliation{Laboratory for Theoretical Cosmology, International Centre of Gravity and Cosmos, Tomsk State University of Control Systems and Radioelectronics, 40 Lenina av., 634050 Tomsk, Russia}

\begin{abstract}
We deal with the $n$-dimensional nonlinear Schr\"{o}dinger equation (NLSE) with a cubic nonlocal nonlinearity and an anti-Hermitian term, which is widely used model for the study of open quantum system. We construct asymptotic solutions to the Cauchy problem for such equation within the formalism of semiclassical approximation based on the Maslov complex germ method. Our solutions are localized in a neighbourhood of few points for every given time, i.e. form some spatial pattern. The localization points move over trajectories that are associated with the dynamics of semiclassical quasiparticles. The Cauchy problem for the original NLSE is reduced to the system of ODEs and auxiliary linear equations. The semiclassical nonlinear evolution operator is derived for the NLSE. The general formalism is applied to the specific one-dimensional NLSE with a periodic trap potential, dipole-dipole interaction, and phenomenological damping. It is shown that the long-range interactions in such model, which are considered through the interaction of quasiparticles in our approach, can lead to drastic changes in the behaviour of our asymptotic solutions. \\
\end{abstract}

%\pacs{52.20.Hv, 52.25.Gg, 52.30.-q}

\keywords{trajectory concentrated states; nonlocal nonlinearity; Maslov complex germ method; dipole-dipole interaction; non-Hermitian operator}

\maketitle

\section{Introduction}
The nonlinear Schr\"{o}dinger equation (NLSE) is a common model of collective excitations in nonlinear media. Its simplest variation with a local cubic nonlinearity and external potential is known as the Gross--Pitaevskii (GP) equation \cite{pitaevskii1999}. If one considers an open system, the NLSE with an anti-Hermitian term comes into play \cite{ashida20}. Such variation of the NLSE is widely used in modelling the propagation of optical pulses in nonlinear media \cite{lederer2008,agrawal12}. When the model includes the source of light and losses \cite{aleksic21,aleksic20}, it deals with a fundamentally open system. The NLSE with an anti-Hermitian term also plays crucial role in modelling the Bose--Einstein condensate (BEC) within the framework of the GP model. The interaction of the BEC with an environment described by the anti-Hermitian terms leads to nontrivial effects in this model such as formation of the vortex lattice \cite{fetter01}. The anti-Hermitian terms are also necessary for the mathematical description of an atom laser \cite{arecchi2000}.

The form of nonlinear terms in the NLSE varies greatly depending on specific physical model. In most cases, the cubic nonlinearity is considered as the simplest physically motivated nonlinearity. In the GP model, such nonlinearity describes the effective mean field of interpaticle interaction. The optics related models such as, e.g., the Haus equation \cite{haus1984} often account for the Kerr effect through the nonlinear term. However, even the simplest cubic nonlinearity leads to quite complex models from the mathematical point of view when one deals with the nonlocal nonlinearity. Returning to the GP model, the nonlocality is motivated by the consideration of long-range interactions such as the dipole-dipole interaction \cite{baranov2008,malomed2009,klaus2022,zhao2021}. In the Haus-like models, where both of the independent variables are associated with time, the nonlocality describes the memory effect of the medium. Besides the nonlocal Kerr effect, the saturation of the laser medium also can be included into the kernel of the nonlocal nonlinearity within the framework of such models \cite{nizette21}. A number of papers are devoted to the mathematical efforts in consideration of various aspects of nonlocality in the NLSE (see, e.g., \cite{curtis2012,kout2024,breev2022}) as well to the derivation of such nonlocal models \cite{bobmann23,boccato15,benedikter14,pickl11,spohn80,marcucci19}. However, it quite hard to obtain exact mathematical results for such complex equations, especially when the nonlocality and the non-Hermiticity are considered at the same time. Hence, many results rely only on numerical calculations \cite{malomed2009,santos16}.

In order to advance in the problem under consideration, we will apply to the semiclassical approximation. A powerful tool that allows one to deals with such problem is the Maslov complex germ method \cite{Maslov2,BeD2}. In \cite{shapovalov:BTS1}, it was shown that this method can be applied to the nonlocal NLSE of a quite general form. Asymptotics for various specific nonlocal NLSE were obtained using the ideas of the Maslov complex germ method in the past years (see, e.g., \cite{lisok2007,athanas11,Pereskokov2017,pereskokov24}). In \cite{kulagin2024}, we have shown that the approach \cite{shapovalov:BTS1} can be generalized to the nonlocal NLSE with an anti-Hermitian term. That approach allows one to construct the asymptotic solutions to the Cauchy problem localized in a neighbourhood of one point moving along the trajectory determined by "classical"\, equations. The limitation of such approach with respect to the physics is that it really deals only with weakly nonlocal effects since the respective asymptotic solutions have trivial geometry. On the other hand, the attractive feature of the nonlocal models is the possibility to consider long-range interactions that can lead to nontrivial spatial patterns \cite{shi2023,saito16}. Thus, we come to the even more complex problem of constructing asymptotic solutions to the nonlocal NLSE with an anti-Hermitian part that effectively depend on the behaviour of the nonlinearity kernel on its whole support rather than the neighbourhood of a center point, i.e. account for the long-range interactions. It turned out to be possible within the framework of the Maslov complex germ based approach if we introduce the so-called semiclassical quasiparticles similar to the ones in \cite{kulshap24} where we dealt with the classical problem of a population dynamics.

The concept of quasiparticles is known in a theory of solitons for nonlinear equations that are exactly integrable in terms of the inverse scattering transform (IST) \cite{zakharov84,dodd82}. The approximate solutions based on this conception can be obtained using the perturbation theory for the equations that are close to the exactly solvable ones with IST \cite{karpman81}, e.g., the $(1+1)$-dimensional Kortewe-de Vries equation, the sine-Gordon equation, the NLSE with the local cubic nonlinearity, and some others. Such soliton solutions are treated as modes of the field excitation. Although our problem is far from the exactly solvable cases, the semiclassical quasiparticles can be treated in a similar way. Usually, the term "quasiparticles"\, implies the components of solution that are distinguishable in either the momentum space (see, e.g. \cite{ribeiro23}) or the coordinate space. The latter one is our case. Note that, for the Haus-like models, where there is no the spatial variables, the quasiparticles can be treated as a train of optical pulses. The distinctive feature of our approach is that the interaction of the semiclassical quasiparticles is ruled by the exact "classical mechanics"\, (dynamical system) rather than by the pertubation of the exactly solvable wave equation.

In this work, we construct the asymptotic solutions to the Cauchy problem for the nonlocal NLSE with an anti-Hermitian term that are semiclassically localized in a neighbourhood of few trajectories associated with the dynamics of quasiparticles. The paper is organized as follows. In Section \ref{sec1}, we give the mathematical statement of the problem under consideration, and clarify the meaning of the semiclassically concentrated states. In Section \ref{sec:quasi}, we introduce the wave functions of quasiparticles that are auxiliary mathematical objects allowing us to construct the approximate solution to the original NLSE. The moments of such wave functions are defined in Section \ref{sec:mom}. The solutions to the dynamical system describing these moments is a key elements of our approach. In Section \ref{sec:cauchy}, we pose the Cauchy problem for the equations associated with the NLSE and give its relation to the original problem. Section \ref{sec:alnlse} is devoted to the reduction of the original complex nonlinear problem to the linear ones within the framework of our quasiparticle formalism. The approximate evolution operator for the original NLSE is constructed. In Section \ref{sec:ex}, we provide the general formalism with a physically motivated example. The one-dimensional NLSE with dipole-dipole interaction, optical-lattice potential, and phenomenological damping is considered. It is shown that the non-perturbative interaction of semiclassical quasiparticles plays a crucial role in the behaviour of the solution to the NLSE. In Section \ref{sec:con}, we conclude with some remarks.

\section{Nonlocal NLSE with an anti-Hermitian term. Classical equations}
\label{sec1}

We deal with a quite general form of the non-Hermitian nonlocal NLSE that reads as follows:
\begin{equation}
\begin{array}{l}
\bigg\{ -i\hbar\pa_t + H(\hat{z},t)[\Psi]-i\hbar \Lambda \breve{H}(\hat{z},t)[\Psi] \bigg\}\Psi(\vec{x},t)=0, \cr \cr
H(\hat{z},t)[\Psi]=V(\hat{z},t)+\varkappa\dil_{{\mathbb{R}}^n}d\vec{y}\,\Psi^{*}(\vec{y},t)W(\hat{z},\hat{w},t)\Psi(\vec{y},t), \cr
\breve{H}(\hat{z},t)[\Psi]=\breve{V}(\hat{z},t)+\varkappa\dil_{{\mathbb{R}}^n}d\vec{y}\,\Psi^{*}(\vec{y},t)\breve{W}(\hat{z},\hat{w},t)\Psi(\vec{y},t).
\end{array}
\label{hartree1}
\end{equation}
Here, $\vec{x}\in {\mathbb{R}}^n$,$\hat{\vec{p}}_x=-i\hbar\pa_{\vec{x}}$, $\hat{z}=(\hat{\vec{p}}_x,\vec{x})$, and $\hat{w}=(\hat{\vec{p}}_y,\vec{y})$. As usual for a semiclassical formalism, an operator of the equation is defined in terms of pseudo-differential operators \cite{Maslov1,maslov81}. So, the operators $V(\hat{z},t)$, $\breve{V}(\hat{z},t)$, $W(\hat{z},\hat{w},t)$, $\breve{W}(\hat{z},\hat{w},t)$ belong to the set ${\mathcal{A}}^t_{\hbar}$ of pseudo-differential operators with smooth symbols growing not faster than polynomial (some related properties and formal definitions are given in Appendix \ref{app0}). Since the equation \eqref{hartree1} is given in the coordinate representation, the operator $\vec{x}$ is the operator of multiplication by $\vec{x}$. Hereinafter, we put a right arrow over and only over the $n$-dimensional vectors. The nonlocal interactions of atoms in BEC usually lead to the kernels $W$ and $\breve{W}$ with "almost compact support", while the saturation of the medium in the optics related models yield the step-like contribution to the $\breve{W}$. Due to the wide variety of kernels in physics, the application of our formalism to the specific models can benefit greatly from the generality of our problem statement.

\begin{defin}
The function $\Psi(\vec{x},t,\hbar)$ belongs to the class ${\mathcal{T}}^t_\hbar\left(\left\{Z_s(t),\mu_s(t)\right\}_{s=1}^{K}\right)$ of functions semiclassically concentrated on trajectories $z=Z_s(t)$ with weights $\mu_s(t)$, $s=\overline{1,K}$, if for any operator $\hat{A}\in{\mathcal{A}}_{\hbar}^{t}$ with a Weyl symbol $A(z,t,\hbar)$ the followings holds:
\begin{equation}
\begin{gathered}
\lim\limits_{\hbar\to 0}\langle\Psi | \hat{A} | \Psi \rangle (t,\hbar)=\sum_{s=1}^{K}\mu_s(t) A(Z_s(t),t,0),\\ \langle\Psi | \hat{A} | \Psi \rangle (t,\hbar)=\dil_{{\mathbb{R}}^n}\Psi^*(\vec{x},t,\hbar)\hat{A} \Psi(\vec{x},t,\hbar)d\vec{x}.
\end{gathered}
\label{def1a}
\end{equation}
\end{defin}

Hereinafter, $s=\overline{1,K}$ stands for $s=1,2,...,K$, where $K\in {\mathbb{N}}$ is interpreted as a total number of quasiparticles. The $2n$-tuple vector $Z_s(t)=\left(\vec{P}_s(t),\vec{X}_s(t)\right)$ can be treated as the phase coordinate of the $s$-th quasiparticle, and $\mu_s(t)$ corresponds to a quantity of matter related to the $s$-th quasiparticle (a "mass"\, of the quasiparticle). We term the weight functions $\mu_s(t)$ as "masses"\, of quasiparticles. Note that $\mu_s(t)$ is not direct analog for a mass of a classical particle since it is not a measure of inertia of the $s$-th particle in a common sense (it will be clear in the specific example in Section \ref{sec:ex}). However, there are some reasons to draw the analogy with masses for $\mu_s(t)$ that are given below.

Let $\mu_{\Psi}(t,\hbar)=||\Psi||^2(t,\hbar)$. Note that the following relation holds in the class ${\mathcal{T}}^t_\hbar\left(\left\{Z_s(t),\mu_s^{(0)}(t)\right\}_{s=1}^{K}\right)$:
\begin{equation}
\lim\limits_{\hbar\to 0}\mu_{\Psi}(t,\hbar)=\sum_{s=1}^{K}\mu^{(0)}_s(t).
\label{limsig1}
\end{equation}
This relation is obtained using the substitution of $\hat{A}=1$ into \eqref{def1a}. Hereinafter, we will denote the second functional parameters of the class ${\mathcal{T}}^t_\hbar\left(\left\{Z_s(t),\mu_s^{(0)}(t)\right\}_{s=1}^{K}\right)$ by $\mu^{(0)}_s(t)$ since it corresponds to the zeroth order approximation (with respect of $\hbar$) of the zeroth moments of the quasiparticle wave function, which will be introduces in Section \ref{sec:mom}.

Let us also consider the mean value of the coordinate $\vec{X}_{\Psi}(t,\hbar)=\dac{1}{\mu_{\Psi}(t,\hbar)}\dil_{{\mathbb{R}}^n}\vec{x}|\Psi(\vec{x},t)|^2$. From \eqref{def1a} one readily gets
the following relation:
\begin{equation}
\lim\limits_{\hbar\to 0}\vec{X}_{\Psi}(t,\hbar)=\dac{\sum_{s=1}^{K}\mu^{(0)}_s(t)\vec{X}_s(t)}{\sum_{s=1}^{K}\mu^{(0)}_s(t)}.
\label{limx1}
\end{equation}
We can treat $\lim\limits_{\hbar\to 0}\vec{X}_{\Psi}(t,\hbar)$ as the coordinate of the center of mass of the system within the semiclassical approximation and $\lim\limits_{\hbar\to 0}\mu_{\Psi}(t,\hbar)$ as its mass. Then, the equations \eqref{limsig1} and \eqref{limx1} correspond to the well-known laws for the mass of a classical system and the position of its center of mass, respectively, where $\mu^{(0)}_s(t)$ act exactly as masses of the quasiparticles that constitute such system.

Now we go on to the derivation of the equations that determine the parameters of the class ${\mathcal{T}}^t_\hbar\left(\left\{Z_s(t),\mu_s^{(0)}(t)\right\}_{s=1}^{K}\right)$ on solutions of \eqref{hartree1}.
The equation for $\mu_{\Psi}(t,\hbar)$ can be readily obtained using the direct substitution of $\pa_t\Psi(\vec{x},t)$ from the original equation \eqref{hartree1} into $\dot{\mu}_{\Psi}(t,\hbar)$ and reads as follows:
\begin{equation}
\begin{gathered}
\dot{\mu}_{\Psi}(t,\hbar)=-2\Lambda \displaystyle\int\limits_{{\mathbb{R}}^n}
d\vec{x}\,\Psi^{*}(\vec{x},t;\hbar) \breve{H}(\hat{z},t)[\Psi]\Psi(\vec{x},t;\hbar)=
-2\Lambda\langle\Psi|\breve{H}[\Psi]|\Psi \rangle=\\
=-2\Lambda\left(\langle\Psi|\breve{V}(\hat{z},t)|\Psi \rangle+\varkappa \langle\Psi|\dil_{{\mathbb{R}}^n}d\vec{y}\,\Psi^{*}(\vec{y},t)\breve{W}(\hat{z},\hat{w},t)\Psi(\vec{y},t)|\Psi \rangle\right).
\end{gathered}
 \label{sig2}
 \end{equation}

Using \eqref{def1a}, we also obtain
\begin{equation}
\begin{gathered}
\lim\limits_{\hbar\to 0}\langle\Psi|\breve{V}(\hat{z},t)|\Psi \rangle=\sum_{s=1}^{K}\mu^{(0)}_s(t)\breve{V}(Z_s(t),t),\\
\lim\limits_{\hbar\to 0}\langle\Psi|\dil_{{\mathbb{R}}^n}d\vec{y}\,\Psi^{*}(\vec{y},t)\breve{W}(\hat{z},\hat{w},t)\Psi(\vec{y},t)|\Psi \rangle=\sum_{s=1}^{K}\sum_{r=1}^{K}\mu^{(0)}_s(t)\mu^{(0)}_{r}(t)\breve{W}(Z_s(t),Z_{r}(t),t).
\end{gathered}
\label{limop1}
\end{equation}
Then, the equation \eqref{sig2}, in the limit $\hbar\to 0$, can be written as follows:
\begin{equation}
\sum_{s=1}^{K}\dot{\mu}^{(0)}_s(t)=-2\Lambda\sum_{s=1}^{K} \mu^{(0)}_s(t)\left( \breve{V}(Z_s(t),t)+\varkappa \sum_{r=1}^{K}\mu^{(0)}_{r}(t)\breve{W}(Z_s(t),Z_{r}(t),t)\right).
\label{limsig2}
\end{equation}

Now, let $A_{\Psi}(t)=\langle \hat{A} \rangle_{\Psi}=\langle \Psi | \hat{A} | \Psi \rangle$. The exact equation for $A_{\Psi}(t)$ is as follows:
\begin{align}
&\displaystyle\frac{\partial}{\partial t}\langle\hat{A}(t)\rangle_\Psi=\bigg\langle\displaystyle\frac{\partial \hat{A}(t)}{\partial t} \bigg\rangle_\Psi +\frac{i}{\hbar}\big\langle \big[H(\hat{z},t)[\Psi],\hat{A}(t)\big]
 \big\rangle_\Psi-\Lambda \big\langle \big[\breve{H}(\hat{z},t)[\Psi],\hat{A}(t)\big]_{+}
 \big\rangle_\Psi=\cr
 &= \bigg\langle\displaystyle\frac{\partial \hat{A}(t)}{\partial t} \bigg\rangle_\Psi+\dac{i}{\hbar} \big\langle [V(\hat{z},t),A(\hat{z},t)]\big\rangle_\Psi - \Lambda\big\langle [\breve{V}(\hat{z},t),A(\hat{z},t)]_{+}\big\rangle_\Psi + \cr
&+\varkappa \bigg\langle\dil_{{\mathbb{R}}^n}d\vec{y}\, \Psi^{*} \Big(\dac{i}{\hbar}[W(\hat{z},\hat{w},t), A(\hat{z},t)]-\Lambda [\breve{W}(\hat{z},\hat{w},t), A(\hat{z},t)]_{+} \Big)\Psi(\vec{y},t)\bigg\rangle_\Psi.
  \label{mean2}
\end{align}

Let us use the property \eqref{limpo1a} of pseudo-differential operators. Then, for the operator $\hat{A}=A(\hat{z},t)$ with a Weyl symbol $A(z,t)$, in the limit $\hbar\to 0$, we come at the following equation:
\begin{equation}
\begin{gathered}
\dac{d}{dt}\sum_{s=1}^{K}\mu^{(0)}_s(t)A(Z_s(t),t)=\sum_{s=1}^{K}\mu^{(0)}_s(t)\Bigg(\dac{\pa A(z_s,t)}{\pa t}-\left\{V(z_s,t),A(z_s,t)\right\}-2\Lambda \breve{V}(z_s,t)A(z_s,t)+\\
+\varkappa\sum_{r=1}^K \mu^{(0)}_r(t)\Big(-\left\{W(z_s,w_r,t),A(z_s,t)\right\}-2\Lambda\breve{W}(z_s,w_r,t)A(z_s,t)\Big)\Bigg)\Big|_{z_s=Z_s(t),\,w_r=Z_r(t)}.
\end{gathered}
\label{limop2}
\end{equation}
In particular, for $A(z,t)=z$, we have
\begin{equation}
\begin{gathered}
\dac{d}{dt}\sum_{s=1}^{K}\mu^{(0)}_s(t)Z_s(t)=\sum_{s=1}^{K}\mu^{(0)}_s(t)\Bigg(JV_z(Z_s(t),t)-2\Lambda \breve{V}(Z_s(t),t)Z_s(t)+\\
+\varkappa\sum_{r=1}^K \mu^{(0)}_r(t)\Big(J W_z(Z_s(t),Z_r(t),t)-2\Lambda\breve{W}(Z_s(t),Z_r(t),t)Z_s(t)\Big)\Bigg).
\end{gathered}
\label{limop3}
\end{equation}
Note that the system of the first order ordinary differential equations (ODEs) \eqref{limsig2}, \eqref{limop3} is closed.

The system of $(2n+1)$ equations \eqref{limsig2}, \eqref{limop3} admits particular solutions that satisfy the following system of $K(2n+1)$ equations:
\begin{equation}
\begin{gathered}
\dot{\mu}^{(0)}_s(t)=-2\Lambda\mu^{(0)}_s(t)\left( \breve{V}(Z_s(t),t)+\varkappa \sum_{r=1}^{K}\mu^{(0)}_{r}(t)\breve{W}(Z_s(t),Z_{r}(t),t)\right), \\
\dot{Z}_s(t)=JV_z(Z_s(t),t)+\varkappa\sum_{r=1}^K \mu^{(0)}_r(t)J W_z(Z_s(t),Z_r(t)),\quad s=\overline{1,K}.
\end{gathered}
\label{hes1}
\end{equation}
We will try an asymptotic solution to the equation \eqref{hartree1} in the class ${\mathcal{T}}^t_\hbar\left(\left\{Z_s(t),\mu_s^{(0)}(t)\right\}_{s=1}^{K}\right)$ where the functional parameters $Z_s(t)$, $\mu^{(0)}_s(t)$, $s=\overline{1,K}$, satisfy the system of "classical"\, equations \eqref{hes1}. We term the system \eqref{hes1} as the zeroth order $K$-particle Hamilton-Ehrenfest system by analogy with \cite{shapovalov:BTS1}.

\section{Wave functions of quasiparticles}
\label{sec:quasi}
%Для построения асимптотических решений $\Psi(\vec{x},t,\hbar)$ введем так называемый "квазиклассический анзац".
Let us introduce the family of classes of trajectory concentrated functions ${\mathcal P}_\hbar^t(Z_s(t), S_s(t,\hbar))$ that reads as follows \cite{shapovalov:BTS1}:
\begin{align}
{\mathcal{P}}_\hbar^t(Z_s(t), S_s(t,\hbar))=\bigg\{
\Phi:\Phi(\vec{x},t,\hbar)=\hbar^{-n/4}\cdot\varphi\Big(\frac{\Delta\vec{x}_s}{\sqrt{\hbar}},
t,\hbar\Big) \cdot\exp\Big[\frac{i}{\hbar}\left(S_s(t,\hbar)+\langle \vec{P}_s(t),\Delta \vec{x}_s  \rangle\right) \Big] \bigg\}.
\label{pth1}
\end{align}
Here, $\Phi(\vec{x},t,\hbar)$ is a general element of the class ${\mathcal{P}}_\hbar^t(Z_s(t), S_s(t,\hbar))$; the real functions $Z_s(t)=(\vec{P}_s(t),\vec{X}_s(t))$ and $S_s(t,\hbar)$ are functional parameters of the class ${\mathcal{P}}_\hbar^t(Z_s(t), S_s(t,\hbar))$; $\Delta\vec{x}_s=\vec{x}-\vec{X}_s(t)$; the function $\varphi(\vec{\xi},t,\hbar)$ belongs to the Schwartz space $\mathbb{S}$ with respect to the variables $\vec{\xi}\in{\mathbb{R}}^n$; the functions $Z_s(t), S_s(t,\hbar)$, and $\varphi(\vec{\xi},t,\hbar)$ smoothly depend on $t$ and regularly depend on $\sqrt{\hbar}$ in a neighbourhood of $\hbar=0$. The notations $\langle \cdot,\cdot \rangle$ stands for the Euclidean scalar product of vectors.

The functions $Z_s(t)$ and $\mu^{(0)}_s(t)$, which determine the trajectory and "mass"\, of the $s$-th quasiparticle, will be subjected to the system of equation \eqref{hes1}, %а функцию $S_s(t)$ подчиним по аналогии с классическим действием соотношению
%\begin{equation}
%\dot{S}_s(t)=\langle \vec{P}_s(t),\dot{\vec{X}}_s(t)\rangle - V(t)-\varkappa\mu_s(t)W(t).
%\label{deist1}
%\end{equation}
and the function $S_s(t,\hbar)$ will be defined later.

We will seek for a solution to the equation \eqref{hartree1} in the set of functions that can be presented as follows:
\begin{equation}
\Psi(\vec{x},t,\hbar)=\sum_{s=1}^{K}\Psi_s(\vec{x},t,\hbar), \quad \Psi_s(\vec{x},t,\hbar)\in {\mathcal P}_\hbar^t(Z_s(t), S_s(t,\hbar)).
\label{razqp1}
\end{equation}
Note that the function $\Psi(\vec{x},t,\hbar)$ does not belong to the family of classes ${\mathcal{P}}_\hbar^t$ in the general case $K>1$. In the specific case $K=1$, the formalism that we propose here can be reduced to the one constructed in \cite{kulagin2024} under some simplifying assumptions. The function $\Psi$ of the form \eqref{razqp1} meets the definition \eqref{def1a}. Thus, $\Psi_s$ can be termed as semiclassical wave function of the $s$-th quasiparticle.

One readily gets, that, if the function $\Psi_s(\vec{x},t,\hbar)$ satisfies the system
\begin{equation}
\begin{array}{l}
\bigg\{ -i\hbar\pa_t + H(\hat{z},t)[\sum_{r=1}^{K}\Psi_r]-i\hbar \Lambda \breve{H}(\hat{z},t)[\sum_{r=1}^{K}\Psi_r] \bigg\}\Psi_s(\vec{x},t)=0, \quad s=\overline{1,K},
\end{array}
\label{hartreeqp1}
\end{equation}
the function $\Psi(\vec{x},t,\hbar)$ given by \eqref{razqp1} obeys the original equation \eqref{hartree1}. For brevity, we will denote the class ${\mathcal P}_\hbar^t(Z_s(t), S_s(t,\hbar))$ as ${\mathcal P}_{\hbar}^t(s)$ where it does not cause the confusion.

It was proved in \cite{shapovalov:BTS1,sym2020} that, for functions from the class ${\mathcal P}_{\hbar}^t(s)$ on a finite time interval $t\in[0;T]$, the following asymptotic estimates hold:
\begin{align}
&\{\Delta\hat{z}_s\}^\alpha=\hat{\Or}(\hbar^{|\alpha|/2}),\quad \Delta\hat{z}_s=(\Delta\hat{\vec{p}}_s,\Delta\vec{x}_s),
\label{estim1a}\\
&\langle\Phi_s|\{\Delta\hat{z}_s\}^\alpha|\Phi_s\rangle =\Or(\hbar^{|\alpha|/2}), \quad \Phi_s\in {\mathcal P}_\hbar^t(s).
\label{estim1}
\end{align}
Here, the following notations are used. The estimate $\hat{A}=\hat{\Or}(\hbar^{m})$, $m\geq 0$,
in \eqref{estim1a} means that
 \begin{align}
 &\frac{||\hat{A}\Phi||}{||\Phi||}=\Or(\hbar^{m}),\quad \Phi\in {\mathcal{P}}_\hbar^t|_s;
 \label{estim2}
 \end{align}
 $Z_s(t)=(\vec{P}_s(t), \vec{X}_s(t))$, $\Delta\hat{z}_s=\hat{z}-Z_s(t)=(\Delta\hat{\vec{p}}_s,\Delta\vec{x}_s)$,
$\Delta\hat{\vec{p}}_s=\hat{\vec{p}}-\vec{P}_s(t)$, $\Delta\vec{x}_s=\vec{x}-\vec{X}_s(t)$;
$\{\Delta\hat{z}_s\}^\alpha$ is the operator determined by the Weyl symbol
$(\Delta{z}_s)^\alpha=(z-Z_s(t))^\alpha$. The multi-index $\alpha\in {\mathbb{Z}}^{2n}_{+}$ ($2n$-tuple)
reads $\alpha=(\alpha_1,\alpha_2, \ldots , \alpha_{2n})$; $\alpha_j\in {\mathbb{Z}}^{1}_{+}$, $j=1,2,\ldots , 2n$; $|\alpha|=\alpha_1+\alpha_2+ \ldots  +\alpha_{2n}$. For $v=(v_1,v_2,\ldots, v_{2n})$, we put $v^{\alpha}=v_1^{\alpha_1}v_2^{\alpha_2}\ldots v_{2n}^{\alpha_{2n}}$.

In particular, we have
 \begin{align}
 &\Delta{x}_{s}=\hat{\Or}(\sqrt{\hbar}),\quad \Delta\hat{p}_{s}=\hat{\Or}(\sqrt{\hbar}),\cr
  &-i\hbar\partial/\partial t -\dot{S}_s(t,\hbar)+\langle{\vec{P}}_s(t),\dot{\vec{X}}_s(t)\rangle +\langle\dot{Z}_s(t),J{\Delta\hat{z}}_s \rangle=\hat{\Or}(\hbar).
  \label{estim3}
  \end{align}
 The functions $\Delta^{(\alpha)}_{\Phi_s}(t,\hbar)$ are $|\alpha|$-th order central moments of the function $\Phi_s$.

Hereinafter, all calculations and commentaries are given for $t\in[0;T]$ where $T<\infty$.

\section{Moments of the functions $\Psi_s$}
\label{sec:mom}

Let us introduce the following definition for m-th order central moments of the functions $\Psi_s(\vec{x},t,\hbar)$:
\begin{equation}
\begin{gathered}
\Delta_{s,j_1 j_2 ... j_m}(t,\hbar)[\Psi_s]=\langle \Psi_s|\Delta \hat{z}_{s,j_1}\Delta\hat{z}_{s,j_2}...\Delta\hat{z}_{s,j_m}|\Psi_s\rangle\Big|_{\text{symmetrized over }j_1,...,j_m}.
\end{gathered}
\label{moms1}
\end{equation}
Indices $j_k$, $k=\overline{1,m}$, stand for a number of the element $z\in{\mathbb{R}}^{2n}$. The formula \eqref{moms1} symmetrized over $j_1,...,j_m$ means that $\Delta_{s,j_1 j_2 ... j_m}(t,\hbar)$ is the expectation of the operator with the Weyl symbol $A(z)=\Delta {z}_{s,j_1}\Delta{z}_{s,j_2}...\Delta{z}_{s,j_m}$, $\Delta z_s=z-Z_s(t)$. Thereinafter, the zeroth order moment of the functions $\Psi_s(\vec{x},t,\hbar)$ will be denoted by
\begin{equation}
\mu_s(t,\hbar)[\Psi_s]=||\Psi_s||^2.
\label{moms2}
\end{equation}

It is clear that $\Delta_{s,j_1 j_2 ... j_m}(t,\hbar)[\Psi_s]=\Or(\hbar^{m/2})$ from \eqref{estim1}.
Let the trajectories $\vec{x}=\vec{X}_s(t)$ do not intersect for different $s$, i.e.
\begin{equation}
Z_r(t)\neq Z_s(t), \quad r\neq s, \quad \forall t\in[0,T].
\label{moms3a}
\end{equation}
Under assumption \eqref{moms3a}, the asymptotic expansion for the operator $H(z,t)[\sum_{r=1}^{K}\Psi_r]$ reads
\begin{equation}
\begin{gathered}
H(\hat{z},t)\Big[\sum_{r=1}^{K}\Psi_r\Big]=V(\hat{z},t)+\varkappa \sum_{r=1}^{K}\bigg(W(\hat{z},w,t)\mu_r(t,\hbar)[\Psi_r]+\\
+\sum_{m=1}^{M} \dac{1}{m!}W_{w_{j_1} w_{j_2} ... w_{j_m}}(\hat{z},w,t)\Delta_{r,j_1 j_2 ... j_m}(t,\hbar)[\Psi_r]\bigg)\Big|_{w=Z_r(t)}+\hat{\Or}(\hbar^{(M+1)/2}).
\end{gathered}
\label{moms3}
\end{equation}
Hereinafter, the summation over repeated indices $j_k$ and $i_k$, $k\in {\mathbb{N}}$, is implied. The summation over other repeated indices is not implied until it is explicitly stated. The similar relation holds for $\breve{H}$ up to changes of $H\to \breve{H}$, $V\to\breve{V}$, and $W\to \breve{W}$. Note that we have taken into account in \eqref{moms3} that the following estimate holds under assumption \eqref{moms3a}:
\begin{equation}
\langle \Psi_r | A(\hat{z},t)|\Psi_s\rangle=\Or(\hbar^{\infty}), \quad r\neq s.
\label{moms4}
\end{equation}

In the class ${\mathcal P}_{\hbar}^t(s)$, the moments \eqref{moms1}, \eqref{moms2} also admit the following expansion:
\begin{equation}
\begin{gathered}
\Delta_{s,j_1 j_2 ... j_m}(t,\hbar)[\Psi_s]=\sum_{k=m}^{M}\hbar^{k/2}\Delta^{(k)}_{s,j_1 j_2 ... j_m}(t)[\Psi_s]+\Or(\hbar^{(M+1)/2}),\\
\mu_s(t,\hbar)[\Psi_s]=\sum_{k=0}^{M}\hbar^{k/2}\mu^{(k)}_s(t)[\Psi_s]+\Or(\hbar^{(M+1)/2}).
\end{gathered}
\label{moms5}
\end{equation}
Hence, the equations \eqref{mean2} for the moments of up to the $M$-th order with the accuracy of $\Or(\hbar^{(M+1)/2})$ form a closed system of ODEs. We term this system as the $M$-th order $K$-particle Hamilton--Ehrenfest system for $M\geq 1$ following \cite{shapovalov:BTS1}. The Hamilton--Ehrenfest systems for $M\leq 3$ are derived in Appendix \ref{app1}.

\section{Cauchy problem}
\label{sec:cauchy}
Let us denote
\begin{equation}
\hat{L}\Big[\sum_{r=1}^{K}\Psi_r\Big]=-i\hbar\pa_t+H(\hat{z},t)\Big[\sum_{r=1}^{K}\Psi_r\Big]-i\hbar\Lambda \breve{H}(\hat{z},t)\Big[\sum_{r=1}^{K}\Psi_r\Big].
\label{moms6}
\end{equation}

We pose the Cauchy problem for \eqref{hartree1} as follows:
\begin{equation}
\Psi(\vec{x},t,\hbar)\Big|_{t=0}=\psi(\vec{x},\hbar).
\label{cauchy1}
\end{equation}
Let the initial condition $\psi(\vec{x},t,\hbar)$ meets the condition \eqref{razqp1}, i.e. we have
\begin{equation}
\psi(\vec{x},\hbar)=\sum_{s=1}^{K}\psi_s(\vec{x},\hbar), \qquad \psi_s(\vec{x},\hbar)\in {\mathcal P}_\hbar^0(Z_s(0), S_s(0,\hbar)).
\label{cauchy2}
\end{equation}
Then, the solution to the Cauchy problem for \eqref{hartreeqp1} with the initial conditions
\begin{equation}
\Psi_s(\vec{x},t,\hbar)\Big|_{t=0}=\psi_s(\vec{x},\hbar), \quad s=\overline{1,K},
\label{cauchy3}
\end{equation}
constitutes the solution to the Cauchy problem \eqref{hartree1}, \eqref{cauchy1}, \eqref{cauchy2} following \eqref{razqp1}.

Let $\BFC$ be set of initial conditions for the Hamilton--Ehrenfest system:
\begin{equation}
\BFC=\left(\mu_s(0,\hbar),\left(\left(\Delta_{s,j_1 j_2 ... j_m}(0,\hbar)\right)_{j_k=1}^{2n}\right)_{m=1}^{\infty}\right)_{s=1}^{K}, \quad k=\overline{1,m}.
\label{cauchy4}
\end{equation}
The parameters $\BFC$ enumerate the parametric family of the solutions to the Cauchy problem for the $M$-th order Hamilton--Ehrenfest system. We need those solutions from this family that correspond to the initial condition \eqref{cauchy3}:
\begin{equation}
\BFC\Big[\big(\psi_s\big)_{s=1}^{K}\Big]=\left(\mu_s(0,\hbar)[\psi_s],\left(\left(\Delta_{s,j_1 j_2 ... j_m}(0,\hbar)[\psi_s]\right)_{j_k=1}^{2n}\right)_{m=1}^{\infty}\right)_{s=1}^K, \quad k=\overline{1,m}.
\label{cauchy5}
\end{equation}
Then, if we replace the moments in \eqref{moms3} with the solutions to the Hamilton--Ehrenfest system corresponding to the initial conditions $\BFC$, the resulting operators $H(\hat{z},t,\BFC)$ and $\breve{H}(\hat{z},t,\BFC)$ can be treated as linear ones parameterized by $\BFC$. The linear operators $H(\hat{z},t,\BFC)$ and $\breve{H}(\hat{z},t,\BFC)$ act on the function $\Psi_s$ in the same way as $H(\hat{z},t)\Big[\sum_{r=1}^{K}\Psi_r\Big]$ and $\breve{H}(\hat{z},t)\Big[\sum_{r=1}^{K}\Psi_r\Big]$ if the parameters $\BFC$ obey the following algebraic condition:
\begin{equation}
\BFC=\BFC\Big[\big(\psi_s\big)_{s=1}^{K}\Big].
\label{cauchy6}
\end{equation}
Note that, within the accuracy of $\Or(\hbar^{(M+1)/2})$, we need to determine only a finite set of elements of $\BFC$, i.e. \eqref{cauchy6} is equal to the finite system of algebraic equations.

\section{Associated linear NLSE}
\label{sec:alnlse}

In the class ${\mathcal P}_{\hbar}^t(s)$, in view of \eqref{hes1}, the estimate \eqref{estim3} yields:
\begin{equation}
\begin{gathered}
-i\hbar \pa_t -\dot{S}_s(t)+\langle \vec{P}_s(t),\dot{\vec{X}}_s(t) \rangle +\\
+\big\langle V_z(Z_s(t),t)+\varkappa\sum_{r=1}^K \mu^{(0)}_r(t)W_z(Z_s(t),Z_r(t),t),\Delta \hat{z}_s \big\rangle=\hat{\Or}(\hbar).
\end{gathered}
\label{moms7}
\end{equation}

Expanding operators $H(\hat{z},t,\BFC)$ and $\breve{H}(\hat{z},t,\BFC)$ into an asymptotic series in a neighbourhood of the trajectory $z=Z_s(t)$, we obtain the following estimate in the class ${\mathcal P}_{\hbar}^t(s)$:
\begin{equation}
\begin{gathered}
H(\hat{z},t,\BFC)=\sum_{m=0}^{M} \dac{1}{k!}H_{z_{j_1} ...z_{j_m}}(z,t,\BFC)\Big|_{z=Z_s(t)} \Delta\hat{z}_{s,j_1}...\Delta\hat{z}_{s,j_m}+\hat{\Or}(\hbar^{(M+1)/2}).
\end{gathered}
\label{moms8}
\end{equation}

Using the estimates \eqref{moms8},\eqref{moms3}, and \eqref{moms7} for $M=1$, in the class ${\mathcal P}_{\hbar}^t(s)$, one readily gets that the operator \eqref{moms6} can be written as
\begin{equation}
\hat{L}\Big[\sum_{r=1}^{K}\Psi_r\Big]=\sum_{m=2}^{M}\hat{L}_s^{(m)}\Big[\sum_{r=1}^{K}\Psi_r\Big]+\hat{\Or}(\hbar^{(M+1)/2}), \quad \hat{L}_s^{(m)}\Big[\sum_{r=1}^{K}\Psi_r\Big]=\hat{\Or}(\hbar^{m/2}),
\label{moms9}
\end{equation}
if we put the functional parameter $S_s(t)$ of the class ${\mathcal P}_{\hbar}^t(s)$ as follows:
\begin{equation}
\begin{gathered}
\dot{S}_s(t,\hbar)=\langle \vec{P}_s(t),\dot{\vec{X}}_s(t) \rangle - \Bigg(V(z,t)+\varkappa\sum_{r=1}^K \bigg(\mu_r(t,\hbar)W(z,w)+\\
+\Delta_{r,j}(t,\hbar)W_{w_j}(z,w,t)\bigg)\Big|_{w=Z_r(t)}\Bigg)\bigg|_{z=Z_s(t)},
\end{gathered}
\label{moms10}
\end{equation}
where summation over repeated index $j$ is implied, and the moments $\mu_r(t,\hbar)$, $\Delta_{r,j}(t,\hbar)$ correspond to the initial conditions \eqref{cauchy6}. Note that we replace the operators $\hat{L}_s^{(m)}\Big[\sum_{r=1}^{K}\Psi_r\Big]$ with the linear operators $\hat{L}_s^{(m)}(\BFC)$ under the algebraic condition \eqref{cauchy6} in view of \eqref{moms6}. In \eqref{moms10} and thereinafter, we will omit the argument $\BFC$ if it does not cause the confusion.

Let us give the explicit form of the operators $\hat{L}_s^{(m)}$ in \eqref{moms9} for $m=2$ and $m=3$:
\begin{equation}
\begin{gathered}
\hat{L}_s^{(2)}=-i\hbar\pa_t-\dot{S}_s(t,\hbar)+\vec{P}_{s,j_1}(t)\dot{\vec{X}}_{s,j_1}(t)+\dac{\varkappa}{2}\sum_{r=1}^K \Delta_{r,j_1 j_2}(t,\hbar)W_{|j_1 j_2}(Z_s(t),Z_r(t),t)-\\
-i\hbar\Lambda \breve{V}(Z_s(t),t)-i\hbar\Lambda\varkappa\sum_{r=1}^K \mu_r(t,\hbar)\breve{W}_{|}(Z_s(t),Z_r(t),t)+\\
+V_{j_1}(Z_s(t),t)\Delta \hat{z}_{s,j_1}+\varkappa\sum_{r=1}^K \mu_r(t,\hbar) W_{i_1|}(Z_s(t),Z_r(t),t)\Delta \hat{z}_{s,i_1}+\\
+\varkappa\sum_{r=1}^K \Delta_{r,j_1}(t,\hbar) W_{i_1|j_1}(Z_s(t),Z_r(t),t)\Delta \hat{z}_{s,i_1}+\dac{1}{2}V_{j_1 j_2}(Z_s(t),t)\Delta \hat{z}_{s,j_1}\Delta \hat{z}_{s,j_2}+\\
+\dac{\varkappa}{2}\sum_{r=1}^K \mu_r(t,\hbar) W_{i_1 i_2|}(Z_s(t),Z_r(t),t)\Delta \hat{z}_{s,i_1} \Delta \hat{z}_{s,i_2},
\end{gathered}
\label{moms11a}
\end{equation}

\begin{equation}
\begin{gathered}
\hat{L}_s^{(3)}=\dac{\varkappa}{6}\sum_{r=1}^K \Delta_{r,j_1 j_2 j_3}(t,\hbar)W_{|j_1 j_2 j_3}(Z_s(t),Z_r(t),t)-i\hbar\Lambda\varkappa\sum_{r=1}^K \Delta_{r,j_1}(t,\hbar)\breve{W}_{|j_1}(Z_s(t),Z_r(t),t)+\\
+\dac{\varkappa}{2}\sum_{r=1}^K \Delta_{r,j_1 j_2}(t,\hbar)W_{i_1|j_1 j_2}(Z_s(t),Z_r(t),t)\Delta \hat{z}_{s,i_1}-i\hbar\Lambda \breve{V}_{j_1}(Z_s(t),t) \Delta \hat{z}_{s,j_1}-\\
-i\hbar\Lambda\varkappa\sum_{r=1}^K \mu_{r}(t,\hbar)\breve{W}_{i_1|}(Z_s(t),Z_r(t),t)\Delta \hat{z}_{s,i_1}+\\
+\dac{\varkappa}{2}\sum_{r=1}^K \Delta_{r,j_1}(t,\hbar)W_{i_1 i_2|j_1}(Z_s(t),Z_r(t),t)\Delta \hat{z}_{s,i_1}\Delta \hat{z}_{s,i_2}+\\
+\dac{1}{6}V_{j_1 j_2 j_3}(Z_s(t),t)\Delta \hat{z}_{s,j_1}\Delta \hat{z}_{s,j_2}\Delta \hat{z}_{s,j_3}+\\
+\dac{\varkappa}{6}\sum_{r=1}^K \mu_{r}(t,\hbar)W_{i_1 i_2 i_3|}(Z_s(t),Z_r(t),t)\Delta \hat{z}_{s,i_1}\Delta \hat{z}_{s,i_2}\Delta \hat{z}_{s,i_3}.
\end{gathered}
\label{moms11b}
\end{equation}
Here, the following notations are used:
\begin{equation}
\begin{gathered}
V_{j_1...j_m}(z,t)=\dac{\pa^m}{\pa z_{j_1}...\pa z_{j_m}}V(z,t),\\
W_{i_1...i_k|j_1...j_m}(z,w,t)=\dac{\pa^{k+m}}{\pa z_{i_1}...\pa z_{i_k}\pa w_{j_1}...\pa w_{j_m}}W(z,w,t).
\end{gathered}
\label{moms12}
\end{equation}
Note that the operators $\hat{L}_s^{(m)}$, $m=\overline{0,\infty}$, are not uniquely defined. The specific form of \eqref{moms11a}, \eqref{moms11b} was chosen for reasons of brevity of the corresponding formulae. Moreover, the functions $\mu_s(t,\hbar)$ and $\Delta_{s,j_1...j_m}(t,\hbar)$ in \eqref{moms11a}, \eqref{moms11b} can be replaced with the approximate ones with accuracy of $\Or(\hbar^2)$. If we replace $\mu_s(t,\hbar)$ and $\Delta_{s,j_1...j_m}(t,\hbar)$ with the approximate ones with accuracy of $\Or(\hbar^{3/2})$ in \eqref{moms11a}, then the formulae \eqref{moms11b} will change.

Functions from the class ${\mathcal P}_{\hbar}^t(s)$ admit the following asymptotic expansions:
\begin{equation}
\Psi_s(\vec{x},t,\hbar)=\Psi^{(0)}_s(\vec{x},t,\hbar)+\hbar^{1/2}\Psi^{(1)}_s(\vec{x},t,\hbar)+\hbar^{2}\Psi^{(2)}_s(\vec{x},t,\hbar)+...,
\label{als1}
\end{equation}
where $\Psi^{(k)}_s(\vec{x},t,\hbar)\in {\mathcal P}_\hbar^t(Z_s(t), S_s(t,\hbar))$.

Let the integration constants \eqref{cauchy6} correspond to the initial conditions that are expanded as follows:
\begin{equation}
\begin{gathered}
\psi_s(\vec{x},\hbar)=\psi^{(0)}_s(\vec{x},\hbar)+\hbar^{1/2}\psi^{(1)}_s(\vec{x},\hbar)+\hbar^{1}\psi^{(2)}_s(\vec{x},\hbar)+..., \\ \psi^{(0)}_s(\vec{x},\hbar)\in {\mathcal P}_\hbar^0(Z_s(0), S_s(0,\hbar)).
\end{gathered}
\label{als2}
\end{equation}
Then, the terms of the asymptotic sequence in \eqref{als1} can be found from the following system of equations:
\begin{equation}
\begin{gathered}
\hat{L}_s^{(2)}(\BFC)\Psi^{(0)}_s(\vec{x},t,\hbar)=0, \\
\hat{L}_s^{(2)}(\BFC)\Psi^{(1)}_s(\vec{x},t,\hbar)=-\hbar^{-1/2}\hat{L}_s^{(3)}(\BFC)\Psi^{(0)}_s(\vec{x},t,\hbar),\\
\hat{L}_s^{(2)}(\BFC)\Psi^{(2)}_s(\vec{x},t,\hbar)=-\hbar^{-1}\hat{L}_s^{(4)}(\BFC)\Psi^{(0)}_s(\vec{x},t,\hbar)-\hbar^{-1/2}\hat{L}^{(3)}_s(\BFC)\Psi^{(1)}_s(\vec{x},t,\hbar),\\
...
\end{gathered}
\label{als3}
\end{equation}

The equation
\begin{equation}
\hat{L}^{(2)}_s(\BFC)\Phi_s(\vec{x},t,\hbar,\BFC)=0,
\label{als4}
\end{equation}
where $\hat{L}^{(2)}(\BFC)$ is given by \eqref{moms11a} is termed as the associated linear Schrodinger equation (ALSE). Actually, it is the parametric family of equations where $\BFC$ are numeric parameters and $s$ refers to the functional parameters $Z_s(t)$ of the class ${\mathcal P}_\hbar^t(Z_s(t), S_s(t,\hbar))$. Note that we subject the function $S_s(t,\hbar)$ to \eqref{moms10}. Hence, it is not an independent functional parameter within our formalism.

The solutions to the ALSE \eqref{als4} constitute the leading terms of asymptotics for $\Psi_s(\vec{x},t,\hbar)\in {\mathcal P}_\hbar^t(Z_s(t), S_s(t,\hbar))$, $s=\overline{1,K}$, if the parameters $\BFC$ meet the algebraic condition \eqref{cauchy6} and $\Phi_s(\vec{x},t,\hbar,\BFC)\Big|_{t=0}=\psi_s(\vec{x},\hbar)$, $s=\overline{1,K}$.

The equation \eqref{als4} can be written as follows:
\begin{equation}
\begin{gathered}
\bigg\{-i\hbar\pa_t+\tilde{H}_s(t,\BFC)+\tilde{H}_{s,i_1}(t,\BFC)\Delta \hat{z}_{s,i_1}+\dac{1}{2}\tilde{H}_{s,i_1 i_2}(t,\BFC)\Delta \hat{z}_{s,i_1}\Delta \hat{z}_{s,i_2}\bigg\}\Phi_s(\vec{x},t,\hbar,\BFC)=0,\\
\tilde{H}_s(t,\BFC)=-\dot{S}_s(t,\hbar)+\vec{P}_{s,j_1}(t)\dot{\vec{X}}_{s,j_1}(t)+\dac{\varkappa}{2}\sum_{r=1}^K \Delta_{r,j_1 j_2}(t,\hbar)W_{|j_1 j_2}(Z_s(t),Z_r(t),t)-\\-i\hbar\Lambda \breve{V}(Z_s(t),t)-i\hbar\Lambda\varkappa\sum_{r=1}^K \mu_r(t,\hbar)\breve{W}_{|}(Z_s(t),Z_r(t),t),\\
\tilde{H}_{s,i_1}(t,\BFC)=V_{i_1}(Z_s(t),t)+\varkappa\sum_{r=1}^K \mu_r(t,\hbar) W_{i_1|}(Z_s(t),Z_r(t),t)+\\+\varkappa\sum_{r=1}^K \Delta_{r,j_1}(t,\hbar) W_{i_1|j_1}(Z_s(t),Z_r(t),t),\\
\tilde{H}_{s,i_1 i_2}(t,\BFC)=V_{i_1 i_2}(Z_s(t),t)+\varkappa\sum_{r=1}^K \mu_r(t,\hbar) W_{i_1 i_2|}(Z_s(t),Z_r(t),t).
\end{gathered}
\label{green1}
\end{equation}

If $\det\bigg(\Big(\tilde{H}_{s,i_1 i_2}(t,\BFC)\Big)_{i_1,i_2=1}^{n}\bigg)\neq 0$, the Green functions for \eqref{green1} reads
\begin{equation}
\begin{gathered}
G_s(\vec{x},\vec{y},t,\hbar,{\bf C})=\displaystyle\frac{1}{\sqrt{\det\big(-2\pi i \hbar M_{s,3}(t)\big)}} \exp\Bigg\{\displaystyle\frac{i}{\hbar}\bigg[\displaystyle\int\limits_0^t \Big( \langle \vec{P}_s(\tau),\dot{\vec{X}}_s(\tau) \rangle-\tilde{H}_s(\tau,{\bf C}) \Big)d\tau+ \\
+\langle \vec{P}_s(t), \Delta \vec{x}_s \rangle - \langle \vec{P}_s(0), \Delta \vec{y}_s \rangle - \displaystyle\frac{1}{2} \langle \Delta \vec{x}_s, M_{s,3}^{-1}(t,\BFC) M_{s,1}(t,\BFC)\Delta \vec{x}_s \rangle +\\
+\langle \Delta \vec{x}_s, M_{s,3}^{-1}(t,\BFC) \Delta \vec{y}_s \rangle - \displaystyle\frac{1}{2} \langle \Delta \vec{y}_s, M_{s,4}(t,\BFC)M_{s,3}^{-1}(t,\BFC)\Delta \vec{y}_s \rangle \bigg] \Bigg\},
\end{gathered}
\label{ur7}
\end{equation}
where $\Delta \vec{y}_s=\vec{y}-\vec{X}_s(0)$ and $2n\times 2n$ block matrices $M_s=\begin{pmatrix} M_{s,1} && -M_{s,3} \cr -M_{s,2} && M_{s,4} \end{pmatrix}$ are solutions to the Cauchy problem
\begin{equation}
\dot{M}_{s,i_1 i_2}(t,\BFC)=-M_{s,i_1 j_1}(t,\BFC)\tilde{H}_{s,j_1 j_2}\big(t,\BFC\big)J_{j_2 i_2}, \qquad M_{s,i_1 i_2}(0)=\delta_{i_1 i_2}.
\label{ur5}
\end{equation}
Then, $\Psi_s^{(0)}$ is given by
\begin{equation}
\Psi_s^{(0)}(\vec{x},t,\hbar)=\dil_{{\mathbb{R}}^n}G_s\Big(\vec{x},\vec{y},t,\hbar,{\bf C}\Big[\big(\psi_s\big)_{s=1}^{K}\Big]\Big)\psi_s^{(0)}(\vec{y},\hbar)d\vec{y}.
\label{green3}
\end{equation}
Note that ${\bf C}\Big[\big(\psi_s\big)_{s=1}^{K}\Big]$ can not be replaced with ${\bf C}\Big[\big(\psi_s^{(0)}\big)_{s=1}^{K}\Big]$ in \eqref{green3} if $\psi_s^{(k)}\neq 0$, $k\geq 1$.
%Let us define the decomposition operator $\hat{D}_s$ on the class of functions given by \eqref{cauchy2} that reads
%\begin{equation}
%\hat{D}_s\psi=\psi_s.
%\label{green4}
%\end{equation}
%Then, the semiclassical nonlinear evolution operator $\hat{U}(t)$ on ${\mathcal P}_{\hbar}^t|_s$ is given by
%\begin{equation}
%\begin{gathered}
%\Psi^{(0)}(\vec{x},t,\hbar)=\hat{U}(t)\psi(\vec{x},\hbar)=\sum_{s=1}^{K}\dil_{{\mathbb{R}}^n}G_s\Big(\vec{x},\vec{y},t,\hbar,{\bf C}\Big[\big(\hat{D}_r\psi\big)_{r=1}^{K}\Big]\Big)\hat{D}_s\psi(\vec{y},\hbar)d\vec{y},
%\end{gathered}
%\label{green5}
%\end{equation}

Actually, we can solve all equations in \eqref{als3} using the Green function \eqref{ur7} and Duhamel`s principle. For example, $\Psi^{(1)}(\vec{x},t,\hbar)$ reads
\begin{equation}
\begin{gathered}
\Psi_s^{(1)}(\vec{x},t,\hbar)=\dil_{{\mathbb{R}}^n} G_{s}\Big(\vec{x},\vec{y},t,\hbar,{\bf C}\Big[\big(\hat{D}_r\psi\big)_{r=1}^{K}\Big]\Big)\psi_s^{(1)}(\vec{y},\hbar)dy+ \\
+\int_{0}^{t}\dac{d\tau}{i\hbar^{3/2}} \dil_{{\mathbb{R}}^n} G_{s}\Big(\vec{x},\vec{y},t-\tau,\hbar,{\bf C}\Big[\big(\hat{D}_r\psi\big)_{r=1}^{K}\Big]\Big)\hat{L}_{s}^{(3)}\Big(\tau,{\bf C}\Big[\big(\hat{D}_r\psi\big)_{r=1}^{K}\Big]\Big)\Psi^{(0)}_s(\vec{y},\tau,\hbar)d\vec{y}.
\end{gathered}
\label{greend1}
\end{equation}
Note that one can put $\psi_s^{(0)}(\vec{y},\hbar)=\psi_s(\vec{y},\hbar)$ in the expansion \eqref{als2} for simplicity of notations. In that case, the first term disappears in \eqref{greend1}, and the relation \eqref{green3} can be written via the decomposition operator $\hat{D}_s$, which is given by $\hat{D}_s\psi(\vec{x},\hbar)=\psi_s(\vec{x},\hbar)$, as follows:
\begin{equation}
\begin{gathered}
\Psi^{(0)}(\vec{x},t,\hbar)=\hat{U}(t)\psi(\vec{x},\hbar)=\sum_{s=1}^{K}\dil_{{\mathbb{R}}^n}G_s\Big(\vec{x},\vec{y},t,\hbar,{\bf C}\Big[\big(\hat{D}_r\psi\big)_{r=1}^{K}\Big]\Big)\hat{D}_s\psi(\vec{y},\hbar)d\vec{y},
\end{gathered}
\label{green5}
\end{equation}

Thus, \eqref{green5} defines the semiclassical nonlinear evolution operator $\hat{U}(t)$ for \eqref{hartree1}.

\section{Example}
\label{sec:ex}
In this Section, we consider the example of the one-dimensional ($n=1$) nonlocal NLSE with an anti-Hermitian term defined by symbols
\begin{equation}
\begin{gathered}
V(z)=\dac{1}{2}p^2+\epsilon \cos x, \qquad W(z,w)=\dac{c^2}{\left((x-y)^2+c^2\right)^{3/2}},
\end{gathered}
\label{exam1}
\end{equation}
where $x,y\in{\mathbb{R}}$.

The function $W(z,w,t)$ from \eqref{exam1} decribes the regularized kernel of the mean field for the dipole-dipole interparticle interaction (see \cite{malomed2009}) where $c>0$ is the regularization parameter. The term $\epsilon \cos x$ describes the optical-lattice potential where $\epsilon$ characterizes its strength.

The one way of description of the Bose--Einstein condensate in open systems within the mean field approximation is the model with the phenomenological damping. Such approach is based on the change of the operator $i\pa_t$ in the NLSE by $(i-\gamma)\pa_t$ where $\gamma>0$ is the rate of the phenomenological damping \cite{choi98}. Under notations \eqref{hartree1}, accurate to the coefficient $\hbar$, it corresponds to the anti-Hermitian term with $\Lambda\approx\gamma$ for small $\gamma$ and its symbols $\breve{V}(z)=V(z)$, $\breve{W}(z,w)=W(z,w)$. Thus, the equation under consideration can be written as follows:

\begin{equation}
\bigg\{ \dac{-i\hbar\pa_t}{1-i\hbar\Lambda} + \dac{1}{2}\hat{p}^2+\epsilon\cos x +\varkappa\dil_{-\infty}^{\infty} \dac{c^2}{\big((x-y)^2+c^2\big)^{3/2}}|\Psi(y,t)|^2dy \bigg\}\Psi(x,t)=0.
\label{exameq1}
\end{equation}

Let us pose the initial condition corresponding to two quasiparticles localized in a neighbourhood of points $x=X_1(0)$ and $x=X_2(0)$. For the sake of simplicity, the density profile of both quasiparticles are supposed to be gaussian. The initial pulses of quasiparticles are set equal to zero. Such initial condition reads.
\begin{equation}
\begin{gathered}
\Psi(x,t,\hbar)\Big|_{t=0}=\psi(x,\hbar),\\
\psi(x,\hbar)=\dac{N_1}{\hbar^{1/4}}\exp\Big(-\dac{\left(x-X_1(0)\right)^2}{2\gamma_1^2\hbar}\Big)+\dac{N_2}{\hbar^{1/4}}\exp\Big(-\dac{\left(x-X_2(0)\right)^2}{2\gamma_2^2\hbar}\Big),
\end{gathered}
\label{exam2}
\end{equation}
where the values $N_1$, $N_2$, $X_1(0)$, $X_2(0)$, $\gamma_1$, and $\gamma_2$ are parameters of the initial condition.

The initial condition \eqref{exam2} can be naturally presented in the form \eqref{cauchy2} as follows:
\begin{equation}
\begin{gathered}
\psi_1(x,\hbar)=\dac{N_1}{\hbar^{1/4}}\exp\Big(-\dac{\left(x-X_1(0)\right)^2}{2\gamma_1^2\hbar}\Big),\\
\psi_2(x,\hbar)=\dac{N_2}{\hbar^{1/4}}\exp\Big(-\dac{\left(x-X_2(0)\right)^2}{2\gamma_2^2\hbar}\Big).
\end{gathered}
\label{exam3}
\end{equation}

For the initial conditions \eqref{exam3}, one readily gets the following initial conditions for the two-particle Hamilton--Ehrenfest system:
\begin{equation}
\begin{gathered}
P_s(0)=0, \qquad \mu_s(0,\hbar)=\gamma_s N_s^2 \sqrt{\pi}, \qquad \Delta_j(0,\hbar)=0,\\
\Delta_{s,11}(0,\hbar)=\dac{\hbar}{2\gamma_s^2}, \qquad \Delta_{s,22}(0,\hbar)=\dac{\hbar\gamma_s^2}{2}, \qquad \Delta_{s,12}(0,\hbar)=\Delta_{s,21}(0,\hbar)=0,\\
\Delta_{s,ijk}(0,\hbar)=0, \qquad i,j,k=1,2, \qquad s=1,2.
\end{gathered}
\label{exam4}
\end{equation}
The parameters $X_s(0)$, $s=1,2$, are initial conditions for functions $X_s(t)$, respectively. Here, we use the notations as in the main body of the article rather than as in Appendix \ref{app1}, i.e. the number of the quasiparticle is given in a subscript, and the numbers of elements of matrices and vectors follow it after a comma. The vectors $X_s$ and $P_s$ do not have the number of element since they are scalar in the one-dimensional case. The initial conditions \eqref{exam4} satisfy \eqref{hohes7}. Hence, the two-particle Hamilton--Ehrenfest system has the simplified form \eqref{hohes8} in this case.

For the example under consideration, the zeroth order two-particle Hamilton--Ehrenfest system (the system \eqref{hohes4} or \eqref{hes1} in the matrix form) is as follows:
\begin{equation}
\begin{gathered}
\dot{P}_1(t)=\epsilon \sin X_1(t)+\varkappa\mu_2^{(0)}(t)c^2\dac{X_1(t)-X_2(t)}{\left(c^2+\left(X_1(t)-X_2(t)\right)^2\right)^{5/2}},\\
\dot{P}_2(t)=\epsilon \sin X_2(t)-\varkappa\mu_1^{(0)}(t)c^2\dac{X_1(t)-X_2(t)}{\left(c^2+\left(X_1(t)-X_2(t)\right)^2\right)^{5/2}},\\
\dot{X}_1(t)=P_1(t),\\
\dot{X}_2(t)=P_2(t),\\
\dot{\mu}^{(0)}_1(t)=-2\Lambda\mu_1^{(0)}(t)\left(\dac{1}{2}P_1^2(t)+\epsilon\cos X_1(t)+\dac{\varkappa\mu_1^{(0)}(t)}{c}+\dac{\varkappa \mu_2^{(0)}(t)c^2}{\left(c^2+\left(X_1(t)-X_2(t)\right)^2\right)^{3/2}}\right),\\
\dot{\mu}^{(0)}_2(t)=-2\Lambda\mu_2^{(0)}(t)\left(\dac{1}{2}P_2^2(t)+\epsilon\cos X_2(t)+\dac{\varkappa\mu_2^{(0)}(t)}{c}+\dac{\varkappa \mu_1^{(0)}(t)c^2}{\left(c^2+\left(X_1(t)-X_2(t)\right)^2\right)^{3/2}}\right).
\end{gathered}
\label{exam5}
\end{equation}
If the regularization parameter $c$ is sufficiently small in the sense that it has a little effect on terms corresponding to the long-range interaction ($c\ll |X_s|$, $s=1,2$, in them), then the system \eqref{exam5} can be reduced to the form
\begin{equation}
\begin{gathered}
\dot{P}_1(t)=\epsilon \sin X_1(t)+\varkappa\mu_2^{(0)}(t)c^2\dac{X_1(t)-X_2(t)}{\left|X_1(t)-X_2(t)\right|^5},\\
\dot{P}_2(t)=\epsilon \sin X_2(t)-\varkappa\mu_1^{(0)}(t)c^2\dac{X_1(t)-X_2(t)}{\left|X_1(t)-X_2(t)\right|^5},\\
\dot{X}_1(t)=P_1(t),\\
\dot{X}_2(t)=P_2(t),\\
\dot{\mu}^{(0)}_1(t)=-2\Lambda\mu_1^{(0)}(t)\left(\dac{1}{2}P_1^2(t)+\epsilon\cos X_1(t)+\dac{\varkappa\mu_1^{(0)}}{c}+\dac{\varkappa \mu_2^{(0)}(t)c^2}{\left|X_1(t)-X_2(t)\right|^{3}}\right),\\
\dot{\mu}^{(0)}_2(t)=-2\Lambda\mu_2^{(0)}(t)\left(\dac{1}{2}P_2^2(t)+\epsilon\cos X_2(t)+\dac{\varkappa\mu_2^{(0)}}{c}+\dac{\varkappa \mu_1^{(0)}(t)c^2}{\left|X_1(t)-X_2(t)\right|^{3}}\right).
\end{gathered}
\label{exam6}
\end{equation}
Let us write all coefficients under notations \eqref{hohes3} for equations of the third order two-particle Hamilton--Ehrenfest system \eqref{hohes4}, \eqref{hohes8} in an explicit form. We will omit the multiplier $\varkappa$ in coefficients \eqref{hohes3} (as if it is equal to 1). For terms corresponding to the long-range interaction we give the exact expression as well as the approximate one (after symbol $\approx$) for a small regularization parameter.
\begin{equation}
\begin{gathered}
V^s=\dac{1}{2}P_s^2(t)+\epsilon\cos X_s(t);\\
V_1^s=P_s(t), \qquad V_2^s=-\epsilon \sin X_s(t);\\
V_{11}^s=1,\qquad V_{12}^s=V_{21}^s=0, \qquad V_{22}^s=-\epsilon \cos X_s(t); \\
V_{abc}^s=\left\{\begin{array}{l}\epsilon \sin X_s(t), \quad a=b=c=2, \cr
0, \quad \text{otherwise};\end{array}\right.\\
W_{|}^{11}=W_{|}^{22}=\dac{1}{c}, \qquad W_{|}^{12}=W_{|}^{21}=\dac{c^2}{\left(c^2+R_{12}^2(t)\right)^{3/2}}\approx\dac{c^2}{|R_{12}(t)|^3};\\
W_{2|}^{11}=W_{2|}^{22}=W_{|2}^{22}=W_{|2}^{11}=0, \\
W_{2|}^{12}=-W_{2|}^{21}=-W_{|2}^{12}=W_{|2}^{21}=\dac{-3c^2R_{12}(t)}{\left(c^2+R_{12}^2(t)\right)^{5/2}}\approx\dac{-3c^2R_{12}(t)}{|R_{12}(t)|^5};\\
W_{22|}^{11}=W_{22|}^{22}=W_{|22}^{11}=W_{|22}^{22}=-W_{2|2}^{11}=-W_{2|2}^{22}=\dac{-3}{c^3}, \\
W_{22|}^{12}=W_{22|}^{21}=W_{|22}^{12}=W_{|22}^{21}-=W_{2|2}^{12}=-W_{2|2}^{21}=\dac{3c^2\left(4R_{12}^2(t)-c^2\right)}{\left(c^2+R_{12}^2(t)\right)^{7/2}}\approx \dac{12c^2}{|R_{12}(t)|^5};\\
W_{222|}^{11}=W_{222|}^{22}=W_{22|2}^{11}=W_{22|2}^{22}=W_{2|22}^{11}=W_{2|22}^{22}=W_{|222}^{11}=W_{|222}^{22}=0,\\
W_{222|}^{12}=-W_{222|}^{21}=-W_{22|2}^{12}=W_{22|2}^{21}=W_{2|22}^{12}=-W_{2|22}^{21}=-W_{|222}^{12}=W_{|222}^{21}=\\
=\dac{15c^2\left(3c^2-4R_{12}^2(t)\right)R_{12}(t)}{\left(c^2+R_{12}^2(t)\right)^{9/2}}\approx \dac{-60c^2R_{12}(t)}{|R_{12}(t)|^7}.
\end{gathered}
\label{exam7}
\end{equation}
Here, we denoted $R_{12}(t)=X_1(t)-X_2(t)$, $R_{12}^2(t)=\left(X_1(t)-X_2(t)\right)^2$. Also, it is implied that all derivative of $W$ with respect to pulses are identically zero.

The leading term of asymptotics, $\Psi^{(0)}(x,t,\hbar)=\Psi^{(0)}_1(x,t,\hbar)+\Psi^{(0)}_2(x,t,\hbar)$, given by \eqref{green3} reads as follows on solutions to two-particle Hamilton--Ehrenfest system and system \eqref{ur5}:
\begin{equation}
\begin{gathered}
\Psi_s^{(0)}(x,t,\hbar)=\dac{N_s\gamma_s}{\hbar^{1/4}\sqrt{\gamma_s^2 M_{s,4}(t)-iM_{s,3}(t)}}\exp\Bigg\{-\dac{\varpi_s(t)\Delta x_s^2}{2\hbar }+\\
+\dac{i}{\hbar}\bigg[\displaystyle\int\limits_0^t \Big( P_s^2(\tau)-\tilde{H}(\tau) \Big)d\tau+P_s(t)\Delta x_s\bigg] \Bigg\},\\
\varpi_s(t)=\dac{\gamma_s^2-\gamma_s^2 M_{s,4}(t)+i M_{s,3}(t)}{M_{s,3}(t)\left(M_{s,3}(t)+i \gamma_s^2 M_{s,4}(t)\right)}.
\end{gathered}
\label{exam8}
\end{equation}
One readily gets that the function \eqref{exam8} meets the initials condition \eqref{exam3}, the relation $\lim_{t\to 0}\varpi_s(t)=\dac{1}{\gamma_s^2}$ is taken into the consideration
The first correction to the leading term of asymptotics \eqref{exam8} given by \eqref{greend1} reads:
\begin{equation}
\begin{gathered}
\Psi_s^{(1)}(x,t,\hbar)=\dil_{0}^{t}d\tau\dac{N_s\gamma_s\left(a^3 c_0 + a^2(b c_1+c_2) + b^3 c_3+ab(bc_2+3c_3)\right)f}{\hbar^{9/4}\sqrt{i M_{s,3}(t-\tau)}\sqrt{i M_{s,3}(\tau)-\gamma_s^2 M_{s,4}(\tau)}a^{7/2}}\exp\left(\dac{b^2}{2a}\right),\\\\
f=\exp\Bigg\{\dac{i}{\hbar}\bigg[\displaystyle\int\limits_0^\tau \Big( P_s^2(\tilde{\tau})-\tilde{H}(\tilde{\tau}) \Big)d\tilde{\tau} +\displaystyle\int\limits_0^{t-\tau} \Big( \langle  P_s^2(\tilde{\tau})-\tilde{H}(\tilde{\tau}) \Big)d\tilde{\tau}+ \\
+P_s(t-\tau) \Delta \tilde{x}_s - \dac{1}{2} M_{s,3}^{-1}(t-\tau) M_{s,1}(t-\tau)(\Delta \tilde{x}_s)^2-\\
-\dac{M_{s,4}(t-\tau)}{2M_{s,3}(t-\tau)}\left(X_s^2(t-\tau)-X_s^2(\tau)\right)+\dac{X_s(\tau)-X_s(t-\tau)}{M_{s,3}(t-\tau)}\Delta\tilde{x}_s\bigg] \Bigg\},\\\\
a=\dac{\varpi_s(\tau)}{\hbar}+\dac{iM_{s,4}(t-\tau)}{\hbar M_{s,3}(t-\tau)},\\\\
b=\dac{i}{\hbar}\bigg[X_s(t-\tau)-X_s(\tau)+P_s(\tau)+ \dac{\Delta\tilde{x}_s}{M_{s,3}(t-\tau)}\bigg],\\\\
c_3=\dac{1}{6}\left(\mu_1(\tau,\hbar)W_{222|}^{s1}(\tau)+\mu_2(\tau,\hbar)W_{222|}^{s2}(\tau)\right)+\dac{1}{6}V_{222}(\tau),\\\\
c_2=\dac{1}{2}\left( \Delta_{1,2}(\tau,\hbar)W_{22|2}^{s1}(\tau)+\Delta_{2,2}(\tau,\hbar)W_{22|2}^{s2}(\tau) \right),\\\\
c_1=\dac{1}{2}\left( \Delta_{1,22}(\tau,\hbar)W_{2|22}^{s1}(\tau)+\Delta_{2,22}(\tau,\hbar)W_{2|22}^{s2}(\tau) \right)-\\
-i\hbar\Lambda \breve{V}_{2}(\tau)-i\hbar\Lambda\left(\mu_1(\tau,\hbar)\breve{W}_{2|}^{s1}(\tau)+\mu_2(\tau,\hbar)\breve{W}_{2|}^{s2}(\tau)\right)+\hbar\breve{V}_1(\tau)\varpi_s(\tau),\\\\
c_0=-i\hbar\Lambda\left(\Delta_{1,2}(\tau,\hbar)\breve{W}_{|2}^{s1}(\tau)+\Delta_{2,2}(\tau,\hbar)\breve{W}_{|2}^{s2}(\tau)\right),\\\\
\Delta\tilde{x}_s=x-X_s(t-\tau).
\end{gathered}
\label{exam9}
\end{equation}
Here, same as in Appendix \ref{app1}, the coefficient $\varkappa$ is included in functions $W$ and $\breve{W}$. We explicitly write the argument $\tau$ for functions $V$, $\breve{V}$, $W$, and $\breve{W}$ implying that the substitution $t=\tau$, including $x=X_s(\tau)$, $y=X_s(\tau)$, $p=P_s(\tau)$, is made in them

\begin{figure}[h]
\begin{minipage}[b][][b]{1\linewidth}\centering
    \includegraphics[width=16 cm]{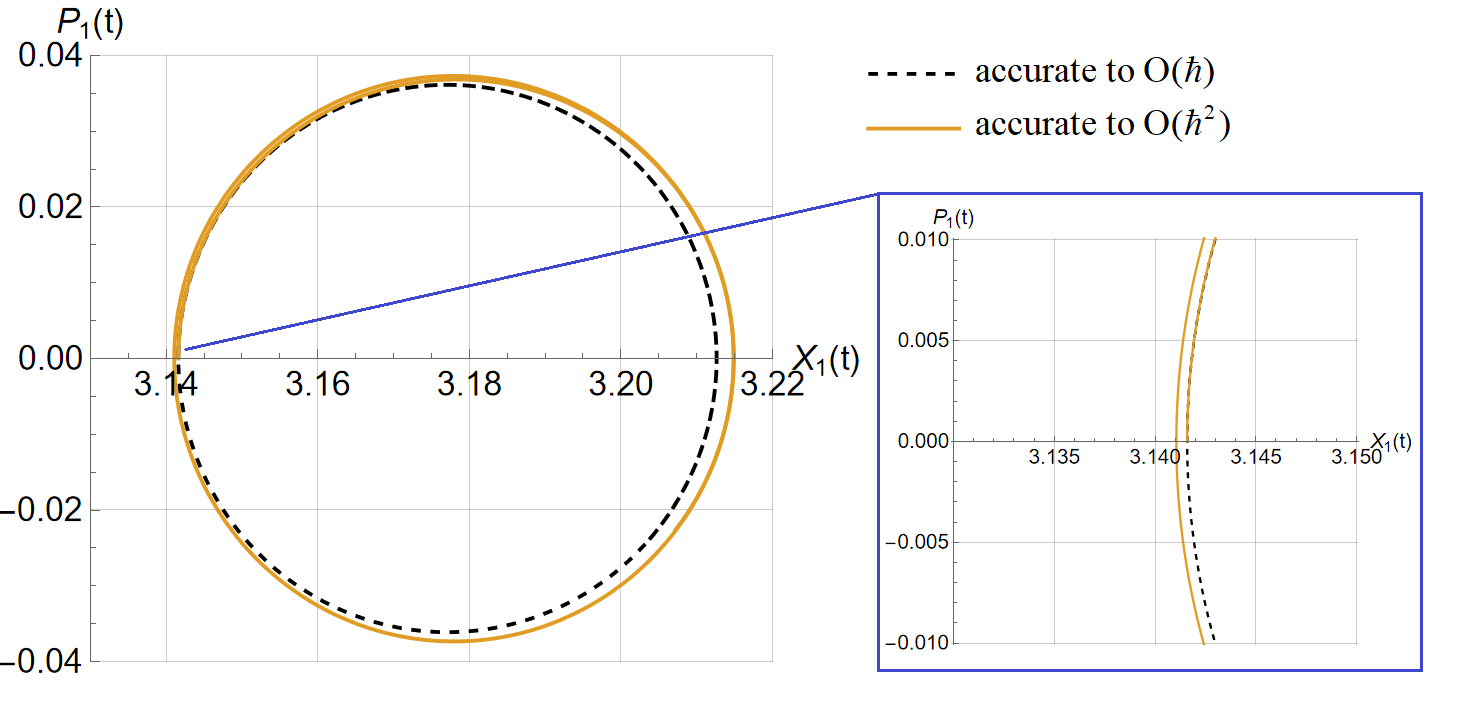}
    \end{minipage}\\
  \caption{Phase trajectory of the first quasiparticle for $\varkappa=2$ and $\Lambda=0$ \label{fig1}}
\end{figure}

Let us consider the initial condition \eqref{exam2}, \eqref{exam3} with parameters that correspond to rest point of the dynamical system for $X_s(t)$, $P_s(t)$, and $\Delta_s^{(2)}(t)$, $s=1,2$, for the linear Hermitian case $\varkappa=\Lambda=0$. Then, we can put
\begin{equation}
X_1(0)=\pi, \qquad X_2(0)=-\pi
\label{exam10a}
\end{equation}
to ensure that $\dot{X}_s(t)=\dot{P}_s(t)=0$, $s=1,2$. The equations for $\Delta_s^{(2)}(t)$ for the linear non-Hermitian case read as follows:
\begin{equation}
\begin{gathered}
\dot{\Delta}_{s,11}^{(2)}(t)=-2V_{22}^s \Delta_{s,12}(t),\\
\dot{\Delta}_{s,22}^{(2)}(t)=2V_{11}^s \Delta_{s,12}(t),\\
\dot{\Delta}_{s,12}^{(2)}(t)=-V_{22}^s \Delta_{s,22}(t)+V_{11}^s \Delta_{s,11}(t).
\end{gathered}
\label{exam10}
\end{equation}
Then, from \eqref{exam4}, \eqref{exam7}, and \eqref{exam10}, one readily gets the following equation for the rest point of \eqref{exam10}:
\begin{equation}
\gamma_s^4=\dac{1}{\epsilon}, \quad s=1,2.
\label{exam11}
\end{equation}
The parameters $N_s$ evidently do not affect the dynamics of the quasiparticles in the linear case. For the sake of simplicity, let us consider the symmetric case $N_1=N_2$. Thus, for the parameters \eqref{exam10a}, \eqref{exam11}, we arrive at the stationary asymptotic solution of the linear Schr\"{o}dinger equation. On the contrary, for the nonlinear case, we obtain the non-stationery solution whose dynamics is caused solely by a nonlinear pertubation of quasiparticles. Let us consider such dynamics starting from the nonlinear Hermitian case ($\varkappa\neq 0$, $\Lambda=0$). Fig. \ref{fig1} shows the phase trajectory of the first ($s=1$) quasiparticle. Since we consider the symmetric case, the trajectory of the second quasiparticle is the same up to the sign of $P_s$ and $X_s$. The Fig. \ref{fig1} is given for $N_1=N_2=1$, $\epsilon=1$, $\hbar=0.1$, $c=3$. The trajectory $Z_{s}(t)$ accurate to $\Or(\hbar)$ is determined by the solutions to the system \eqref{exam5}, while the trajectory accurate to $\Or(\hbar^2)$ is found as $Z_{s}(t,\hbar)=Z_{s}(t)+\dac{\hbar\big(\Delta_{s,1}^{(2)}(t),\Delta_{s,2}^{(2)}(t)\big)}{\mu_s^{(0)}(t)+\hbar\mu_s^{(2)}(t)}$, where $\mu_s(t)=\const$ for $\Lambda=0$.

The solutions of the system \eqref{exam5} yield the closed trajectory as it is clear from Fig. \ref{fig1}. From the linearized form (with respect to the variation of $Z^{(0)}_s(t)$) of the system \eqref{exam5} one readily gets the approximate period of the trajectory that is $T_Z \approx 2\pi \bigg(\epsilon+\varkappa \gamma_1 N_1^2 \sqrt{\pi} c^2\dac{32\pi^2-2c^2}{(c^2+4\pi^2)^{7/2}}\bigg)$ (approximately 6.36 for the values of parameters for Fig. \ref{fig1}). The trajectory becomes nonperiodic if we take into consideration the solutions to the second order two-particle Hamilton--Ehrenfest system, $\Delta_{s,i}^{(2)}(t)$ and $\mu^{(2)}_s(t)$. Although, it is almost periodic for relatively small $t$.

\begin{figure}[h]
\begin{minipage}[b][][b]{1\linewidth}\centering
    \includegraphics[width=16 cm]{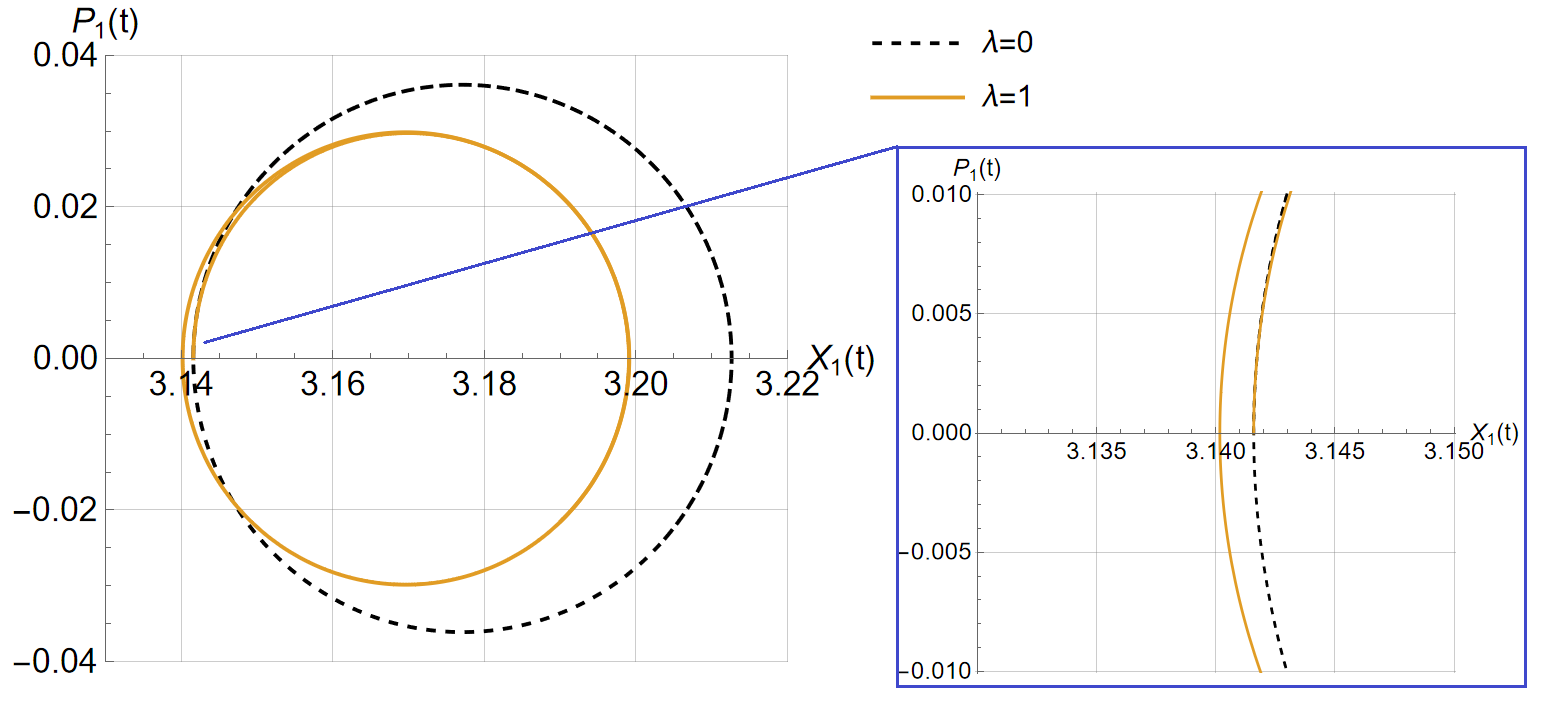}
    \end{minipage}\\
  \caption{Phase trajectory of the first quasiparticle for $\varkappa=2$ and various $\Lambda$ \label{fig2}}
\end{figure}

\begin{figure}[h]
\begin{minipage}[b][][b]{1\linewidth}\centering
    \includegraphics[width=11 cm]{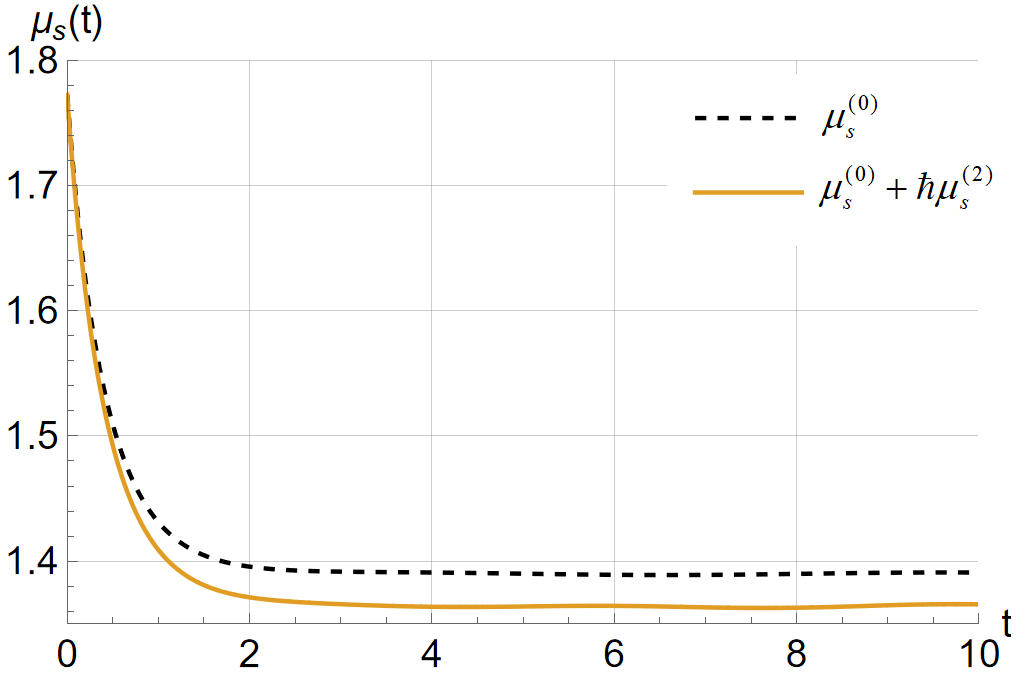}
    \end{minipage}\\
  \caption{Evolution of the mass of a quasiparticle for $\varkappa=2$ and $\Lambda=1$ \label{fig3}}
\end{figure}

\begin{figure}[h]
\begin{minipage}[b][][b]{1\linewidth}\centering
    \includegraphics[width=11 cm]{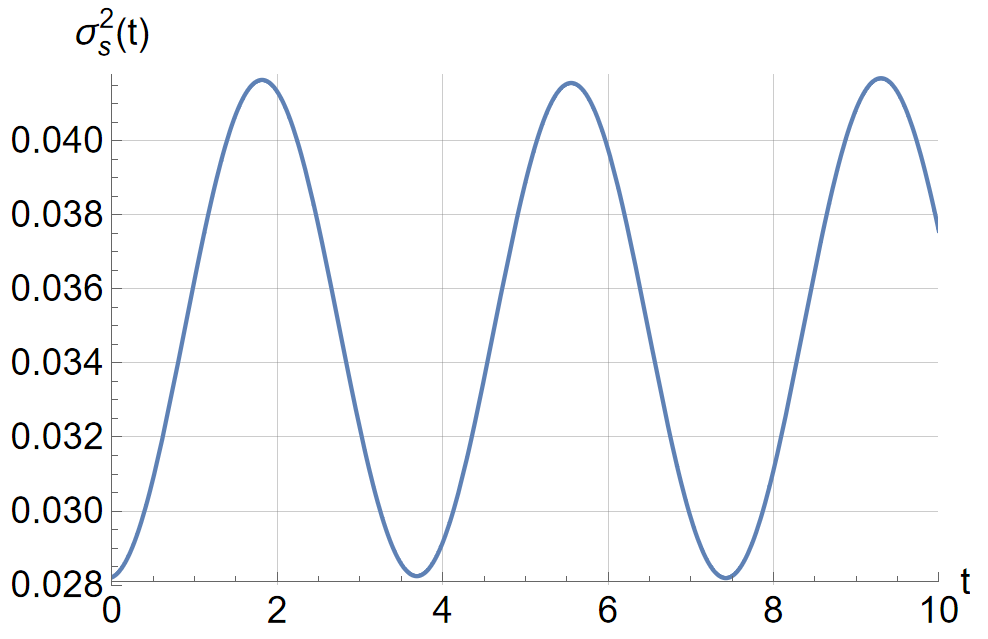}
    \end{minipage}\\
  \caption{Dispersion of the wave function of a quasiparticle for $\varkappa=2$ and $\Lambda=1$ \label{fig4}}
\end{figure}

Now, let us go on to the non-Hermitian case. Fig. \ref{fig2} shows the comparison of trajectories obtained from the system \eqref{exam5} for $\Lambda=0$ and $\Lambda=1$. The trajectory corresponding to the open system ($\Lambda=1$) is evidently nonperiodic due to the transient process. The evolution of quasiparticle masses (they are equal in the symmetric case under consideration) are shown in Fig. \ref{fig3}. Also, the dispersion $\sigma_s^2(t)$ of the wave function of a quasiparticle is given in Fig. \ref{fig4} that reads as follows:
\begin{equation}
\sigma_s^2(t)=\dac{\hbar\cdot\Delta_{s,22}^{(2)}(t)}{\mu_s^{(0)}(t)+\hbar\mu_s^{(2)}(t)}.
\label{disper1}
\end{equation}

%Дисперсия --- multiperiodic???. Dominating oscillations have the period $T_{\sigma^2}\approx 2\pi (2\mu \varkappa)^{-1/2}\bigg(-\dac{12c^2 4\pi^2-3c^4}{(c^2+4\pi^2)^{7/2}}+\dac{3}{c^3}\bigg)^{-1/2}$

\begin{figure}[h]
\begin{minipage}[b][][b]{1\linewidth}\centering
    \includegraphics[width=14 cm]{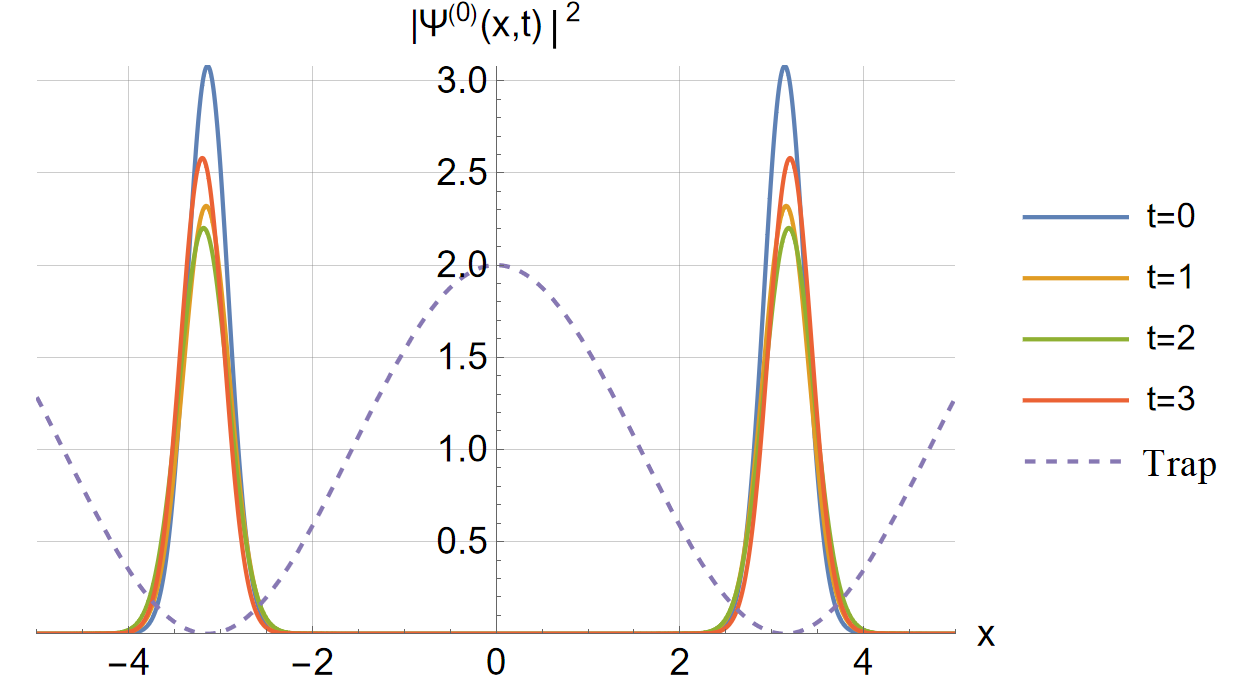}
    \end{minipage}\\
  \caption{Squared absolute value of the asymptotic solution to \eqref{exameq1} for two quasiparticles, $\varkappa=2$, and $\Lambda=1$ \label{fig5}}
\end{figure}

Since the dispersion of the wave functions for each quasiparticle is bounded periodic function, the asymptotic solution to \eqref{exameq1} behave as two interacting and oscillating soliton-like wave functions. The density corresponding to this solution is presented in Fig. \ref{fig5}. This solution is constructed using relation \eqref{exam8}. The contribution of the higher correction \eqref{exam9} turned out to be very small in this specific case (the corrections to Fig. \ref{fig5} are visually indistinguishable). The dashed line in Fig. 5 shows the spatial form of the external trap, $\epsilon\cos x$.

\begin{figure}[h]
\begin{minipage}[b][][b]{1\linewidth}\centering
    \includegraphics[width=11 cm]{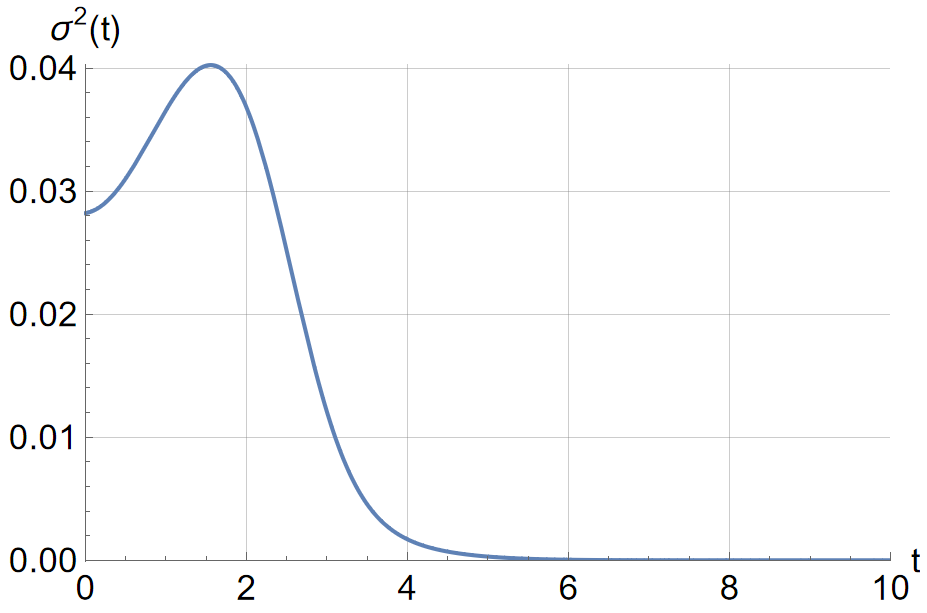}
    \end{minipage}\\
  \caption{Dispersion of the asymptotic solutions corresponding of a single quasiparticle for $\varkappa=2$ and $\Lambda=1$ \label{fig6}}
\end{figure}

Note that in the considered case the interparticle interaction was relatively small in the sense that the trajectories deviate from the stationary point for the small value $d_{12}\ll 2\pi$. It is due to that the effective potential of the dipole-dipole interaction quickly decrease with the increase of range. Thus, one would expect that the solution oscillate near the stationary one if the system is stable. However, let us look on the dispersion of the wave function of a single quasiparticle ($K=1$) that is equal to the asymptotic solution to \eqref{exameq1} in this case. Fig. \ref{fig6} shows one for the same values of parameters except the number of quasiparticles (it is equal to putting $N_2=0$). The dispersion quickly tends to zero with a time, i.e. the solution experiences a collapse. It means that we observed the soliton-like behaviour of the ensemble of interacting quasiparticles for the case $K=2$ and the interaction between quasiparticles that is quite small in the mentioned sense is crucial for such behaviour. Evidently, the common perturbative calculations with respect to the parameter of the long-range interaction can not catch such effect. It is worth to mention that the dispersion does not tend to zero so quickly for a single quasiparticle in the case of a closed system ($\Lambda=0$).
%\\{\bf Тут (или в заключении) еще хотел выразить мысль, что это подчеркивает важность учета нетривиальности области локализации решения. Что сложно заранее сказать, можно ли рассматривать открытую квантовую систему приближенно как две отдельные квантовые системы со слабым взаимодействием между ними, так как вот такие эффекты могут вылезть. А в нашем методе мы существенным образом учитываем, что это одна система, так как мы делаем пертурбативную квантовую надстройку над классической механикой двух квазичастиц. Не знаю, как это лучше сформулировать.\\
%
%Волновые функции квазичастиц --- это способ описания решения со сложной геометрией. Квазичастицы --- это K классических частиц, объединенных общим фазовым пространством, взаимодействием между которыми в нем определяется самим NLSE.}

%\vspace{2em}
%{\bf ЧТО ОСТАЛОСЬ СДЕЛАТЬ:\\
%1) Ввести всякие аббревиатуры для BEC, HES, leading term of asymptotics и так далее. А может и не надо. Сейчас все используемые аббревиатуры объяснены, а не объясненные - не используются. }

\section{Conclusion}
\label{sec:con}
We have developed the formalism that allows one to construct the asymptotic solution $\Psi(\vec{x},t)$ to the Cauchy problem for the $n$-dimensional NLSE \eqref{hartree1} with a nonlocal nonlinearity and non-Hermitian operator within the semiclassical approximation ($\hbar\to 0$). The solution has the following geometric structure. The main share of matter (if we associate $|\Psi(\vec{x},t)|^2$ with the matter density) is localized in a neighbourhood of a finite number $K$ of points moving along the trajectories that are determined by the auxiliary dynamical systems \eqref{hes1}. This system is treated as equations of "classical mechanics"\, for $K$ quasiparticles with time-dependent weights ("masses"). The time dependent weights ("masses") are governed by the anti-Hermitian terms in the original NLSE \eqref{hartree1}, while the equations for phase trajectories of quasiparticles are generated by the Hermitian terms. We have introduced the wave functions of the quasiparticles, $\Psi_s(\vec{x},t)$, that allow us to explicitly obtain the semiclassical nonlinear evolution operator for \eqref{hartree1}. The construction of such approximate evolution operator relies on solutions to the Cauchy problem for the system of ODEs (the Hamilton--Ehrenfect system) describing the quantum corrections to the "classical equations"\, \eqref{hes1}. Thus, the geometric properties of the asymptotic solutions under consideration are associated with the phase trajectories of the quasiparticles. These quasiparticles are close to the ones in the soliton theory \cite{zakharov84,dodd82} in the sense that they can be treated as modes of the excitations of nonlinear system. However, in our asymptotic expansion, we do not rely on exact solutions to the soliton equations but do on the solutions to the "classical equations"\, for the system with $(2n+1)K$ degrees of freedom and to the system of ODEs for the moments of the wave functions of quasiparticles. It fundamentally differs our approach from the common perturbation theory for solitons \cite{karpman81}. For the single quasiparticle ($K=1$), the approach proposed is reduced to the one presented in \cite{kulagin2024}. If we additionally omit the anti-Hermitian terms in \eqref{hartree1} (put $\Lambda=0$), we come to the method given in \cite{shapovalov:BTS1}. Finally, for the linear case ($\varkappa=0$), our approach become the interpretation of the well-known Maslov complex germ method \cite{Maslov2}. Note that specific case $\Lambda=0$, $\varkappa \neq 0$ also was not considered anywhere for $K>1$.

The approach proposed is illustrated with the physically motivated example in Section \ref{sec:ex}. We have applied our method to the one-dimensional NLSE with the periodic external potential corresponding to the optical-lattice, the dipole-dipole interaction, and the phenomenological damping. Our asymptotic analysis shows that the soliton-like behaviour of the solution with the two-point geometry of localization is conditioned on the mechanism of the interaction between quasiparticles in such model.

The results for the specific NLSE \eqref{exameq1} emphasize the importance of the dynamical system derived in this work within the framework of the semiclassical approximation, since it determines the geometric properties and qualitative behaviour of the solution to the NLSE that describes the open quantum system. The comprehensive analysis of this dynamical system, including the symmetry analysis \cite{obukhov2022,osetrin2021}, can be subject for a separate study. Moreover, it is of interest to apply our formalism to the model equation of the optical pulse propagation in nonlinear media in order to discover the physical effects that can be conditioned on the mechanics of the interacting quasiparticles in such physical interpretation of the NLSE. These are the prospects for the future researches.

\section*{Acknowledgement}

The study is supported by Russian Science Foundation, project no. 23-71-01047, https://rscf.ru/en/project/23-71-01047/.

\appendix

\section{Pseudo-differential operators}
\label{app0}
Let functions $A(\vec{p},\vec{x},t,\hbar)$, $\vec{p},\vec{x}\in{\mathbb{R}}^{n}$ from the class ${\mathcal{S}}_\hbar^t$ satisfy the following conditions for every fixed $t\geq 0$:\\
1) $A(\vec{p},\vec{x},t,\hbar)\in C^{\infty}$ with respect to $\vec{p}$ and $\vec{x}$;\\
2) $A(\vec{p},\vec{x},t,\hbar)$ and all its derivatives grow not faster that polynomials of $|\vec{p}|$ and $|\vec{x}|$ as $|\vec{p}|,|\vec{x}|\to \infty$;\\
3) $A(\vec{p},\vec{x},t,\hbar)$ regularly depends on the parameter $\hbar$ in a neighborhood of $\hbar=0$. \\
We also use the brief notation $A(z,t,\hbar)=A(\vec{p},\vec{x},t,\hbar)$, $z=(\vec{p},\vec{x})\in{\mathbb{R}}^{2n}$.

\begin{defin}
A pseudo-differential Weyl-ordered operator is an operator $\hat{A}=A(\hat{z},t,\hbar)=A(\hat{\vec{p}},\vec{x},t,\hbar)$ that is defined by \cite{Maslov1}
\begin{equation}
A(\hat{\vec{p}},\vec{x},t,\hbar)\Phi(\vec{x},t,\hbar)=\dac{1}{(2\pi\hbar)^n}\dil_{{\mathbb{R}}^{2n}}d\vec{p}d\vec{y} \exp\Big(\dac{i}{\hbar}\langle \vec{p},\vec{x}-\vec{y}\rangle\Big)A\Big(\vec{p},\dac{\vec{x}+\vec{y}}{2},t,\hbar\Big)\Phi(\vec{y},t,\hbar),
\label{pseud1}
\end{equation}
where $A(\vec{p},\vec{x},t,\hbar)\in{\mathcal{S}}_\hbar^t$ and $\Psi(\vec{x},t,\hbar)\in{\mathbb{S}}$ for fixed $t$, $\hbar$. Here, ${\mathbb{S}}$ is the Schwartz space, and $\langle a, b \rangle=\sum_{i=1}^{n}a_i b_i$ is the Euclidian scalar product.

The function $A(z,t,\hbar)$ in \eqref{pseud1} is termed the Weyl symbol of the operator $\hat{A}=A(\hat{z},t,\hbar)$.

We denote by ${\mathcal{A}}_\hbar^t$ the set of pseudo-differential operators defined above.
\end{defin}

Note that the Weyl ordering is the "symmetric"\, ordering since it leads to the usual symmetrization for differential operators. For example, if $A(p,x,t,\hbar)=px$, $n=1$, the respective Weyl-ordered operator reads $A(\hat{p},x,t,\hbar)=\dac{1}{2}\left(\hat{p}x+x\hat{p}\right)$.

For the semiclassical approximation theory, the following property of pseudo-differential operators is useful. Let the pseudo-differential operators $C(\hat{z},t)$ and $D(\hat{z},t)$ be given by
\begin{equation}
\begin{gathered}
C(\hat{z},t)=\big[A(\hat{z},t),B(\hat{z},t)\big]=A(\hat{z},t)B(\hat{z},t)-B(\hat{z},t)A(\hat{z},t), \\ D(\hat{z},t)=\big[A(\hat{z},t),B(\hat{z},t))\big]_{+}=A(\hat{z},t)B(\hat{z},t)+B(\hat{z},t)A(\hat{z},t),
\end{gathered}
\label{limpo0}
\end{equation}
where $A(\hat{z},t)$ and $B(\hat{z},t)$ are pseudo-differential operators with the Weyl symbols $A(z,t)$ and $B(z,t)$ respectively.

Then, their Weyl symbols $C(z,t)$ and $D(z,t)$ obey the following relations \cite{multiind}:
\begin{equation}
\begin{gathered}
\lim_{\hbar\to 0} \dac{C(z,t)}{i\hbar}=\big\{A(z,t),B(z,t)\big\}, \qquad \lim_{\hbar\to 0} D(z,t)=2A(z,t)B(z,t).
\end{gathered}
\label{limpo1a}
\end{equation}
where $\big\{A(z,t),B(z,t)\big\}=\bigg\langle\displaystyle\frac{\partial A(z,t)}{\partial z},J \frac{\partial B(z,t)}{\partial z}\bigg\rangle$ is the Poisson bracket, $J=\begin{pmatrix} 0 & -{\mathbb{I}}_{n\times n} \cr {\mathbb{I}}_{n\times n} & 0 \end{pmatrix}$.

\section{Higher order Hamilton--Ehrenfest system}
\label{app1}
In this Appendix, for the given $s$-th quasiparticle, let us denote
\begin{equation}
\begin{gathered}
H_{j_1 j_2 ... j_m}=\dac{\pa^m H(z,t)[\Psi]}{\pa z_{j_1} \pa z_{j_2} ... \pa z_{j_m}}\bigg|_{z=Z_s(t)},\\
\breve{H}_{j_1 j_2 ... j_m}=\dac{\pa^m \breve{H}(z,t)[\Psi]}{\pa z_{j_1} \pa z_{j_2} ... \pa z_{j_m}}\bigg|_{z=Z_s(t)},
\end{gathered}
\label{hohes1}
\end{equation}
where functions $H$, $\breve{H}$ are symmetrized with respect to $j_1, j_2, ..., j_m \in {\mathbb{Z}}_+^{2n}$.

In the class ${\mathcal P}_{\hbar}^t(s)$, the equations \eqref{mean2} yield the following accurate to $\Or(\hbar^2)$:
\begin{equation}
\begin{gathered}
{\color{blue}\dot{\mu}_s}=-\Lambda\left( 2\breve{H} \mu_s + 2 \breve{H}_{a} \Delta_{s,a} + \breve{H}_{ab}\Delta_{s,ab}+\dac{1}{3}\breve{H}_{abc}\Delta_{s,abc}\right),\\
{\color{blue}\dot{\Delta}_{s,i}}=J_{ia} H_a \mu_s +J_{ia}H_{ab}\Delta_{s,b}+\dac{1}{2}J_{ia}H_{abc}\Delta_{s,bc}+\dac{1}{6}J_{ia}H_{abcd}\Delta_{s,bcd}-\\
-\Lambda\left(2\breve{H} \Delta_{s,i} + 2\breve{H}_a \Delta_{s,ai}+\breve{H}_{ab}\Delta_{s,abi}\right)-\dot{Z}_{s,i} \mu_s,\\
{\color{blue}\dot{\Delta}_{s,ij}}=2J_{ia}H_a\Delta_{s,j}+2J_{ia}H_{ab}\Delta_{s,bj}+J_{ia}H_{abc}\Delta_{s,bcj}-\\
-\Lambda\left(2\breve{H}\Delta_{s,ij}+2\breve{H}_a \Delta_{s,aij}\right)-2\Delta_{s,i} \dot{Z}_{s,j}.\\
{\color{blue}\dot{\Delta}_{s,ijk}}=3J_{ia}H_a\Delta_{s,jk}+3J_{ia}H_{ab}\Delta_{s,jkb}-2\Lambda \breve{H} \Delta_{s,ijk}-3\Delta_{s,ij}\dot{Z}_{s,k}.
\end{gathered}
\label{hohes2}
\end{equation}
Hereinafter, we imply the symmetrization with respect to indices $i,j,k\in {\mathbb{Z}}_+^{2n}$ in each term and summation over the repeated indices $a,b,c,d\in {\mathbb{Z}}_+^{2n}$. E.g., we have
\begin{equation}
A_{ia} B_{jak}=\dac{1}{3!}\sum_{a=1}^{2n}\Big(A_{ia} B_{jak}+A_{ia} B_{kaj}+A_{ja} B_{iak}+A_{ka} B_{iaj}+A_{ja} B_{kai}+A_{ka} B_{jai}\Big).
\label{symmet1}
\end{equation}
Also, we highlight time derivatives in blue in this Appendix for better readability.

%, а также обозначено
%\begin{equation}
%\begin{gathered}
%H_{j_1 j_2 ... j_m}=\dac{\pa^m H(z,t)[\Psi]}{\pa z_{j_1} \pa z_{j_2} ... \pa z_{j_m}}\bigg|_{z=Z(t)},\\
%\breve{H}_{j_1 j_2 ... j_m}=\dac{\pa^m \breve{H}(z,t)[\Psi]}{\pa z_{j_1} \pa z_{j_2} ... \pa z_{j_m}}\bigg|_{z=Z(t)}, \quad j_1, j_2, ..., j_m \in {\mathbb{Z}}_+^{2n}.
%\end{gathered}
%\label{hohes1}
%\end{equation}

Using expansions \eqref{moms5}, \eqref{moms3}, \eqref{moms8}, and the contracted notations
\begin{equation}
\begin{gathered}
\mu^{s(k)}=\mu^{(k)}_s(t)[\Psi_s], \quad \Delta_{j_1...j_m}^{s(k)}=\Delta_{s,j_1...j_m}^{(k)}(t)[\Psi_s], \quad \sum_r=\sum_{r=1}^{K}, \\
V_{j_1 j_2 ... j_m}^s=\dac{\pa^m V(z,t)}{\pa z_{j_1} \pa z_{j_2} ... \pa z_{j_m}}\bigg|_{z=Z_s(t)},\\
W_{i_1 i_2 ... i_q|j_1 j_2 ... j_r}^{sr}=\varkappa\dac{\pa^{q+r} W(z,w,t)}{\pa z_{i_1} ... \pa z_{i_q}\pa w_{j_1} ... \pa w_{j_r}}\bigg|_{z=Z_s(t),\, w=Z_r(t)},\\
\breve{V}_{j_1 j_2 ... j_m}^s=\dac{\pa^m \breve{V}(z,t)}{\pa z_{j_1} \pa z_{j_2} ... \pa z_{j_m}}\bigg|_{z=Z_s(t)},\\
\breve{W}_{i_1 i_2 ... i_q|j_1 j_2 ... j_r}^{sr}=\varkappa\dac{\pa^{q+r} \breve{W}(z,w,t)}{\pa z_{i_1} ... \pa z_{i_q}\pa w_{j_1} ... \pa w_{j_r}}\bigg|_{z=Z_s(t),\, w=Z_r(t)},
\end{gathered}
\label{hohes3}
\end{equation}
we obtain the $K$-particle Hamilton--Ehrenfest systems of up to the third order inclusive. In equations of this Appendix, we omit the upper indices $s$ and $r$ for functions $V$, $\breve{V}$, $W$, and $\breve{W}$ in order to make clear that there is no summation over them as over the repeated indices (the summation over $r$ will be given explicitly where it is needed). Such contracted notations do not cause confusion since these indices always go in the same order as in \eqref{hohes3}. Nevertheless, the superscripts are convenient for writing these functions explicitly for specific cases as in Section \ref{sec:ex}. If we do not differentiate the function $W$ or $\breve{W}$ with respect to eigther $z$ or $w$ (or both of them), we still keep the delimiter $|$, e.g.
\begin{equation}
\begin{gathered}
W_{|j_1 j_2}=\varkappa\dac{\pa^{2} W(z,w,t)}{\pa w_{j_1} \pa w_{j_2}}\bigg|_{z=Z_s(t),\, w=Z_r(t)},\\
\breve{W}_{|}=\varkappa \breve{W}(z,w,t)\Big|_{z=Z_s(t),\, w=Z_r(t)}.
\end{gathered}
\label{hohes3a}
\end{equation}

The zeroth order $K$-particle Hamilton--Ehrenfest system, which is given in \eqref{hes1}, reads as follows under notations \eqref{hohes3}:
\begin{equation}
\begin{gathered}
{\color{blue}\dot{Z}^{s(0)}_i}=J_{ia} V_a + \sum_r J_{ia} W_{a|} \mu^{r(0)},\\
{\color{blue}\dot{\mu}^{s(0)}}=-\Lambda \Bigg( 2\breve{V} \mu^{s(0)}+ \sum_r \breve{W}_{|} \mu^{r(0)}\mu^{s(0)} \Bigg).
\end{gathered}
\label{hohes4}
\end{equation}

The first order $K$-particle Hamilton--Ehrenfest system (accurate to $\Or(\sqrt{\hbar})$) includes \eqref{hohes4} and the following equations:
\begin{equation}
\begin{gathered}
{\color{blue}\dot{\Delta}^{s(1)}_{i}}=J_{ia} \sum_r W_{a|} \mu^{r(1)}\mu^{s(0)}+J_{ia} \sum_r W_{a|b} \Delta_b^{r(1)}\mu^{s(0)}+J_{ia}V_{ab}\Delta_{a}^{s(1)}+J_{ia}\sum_r W_{ab|}\Delta_{b}^{s(1)}\mu^{r(0)}-\cr
-\Lambda\left( 2\breve{V} \Delta_{i}^{s(1)}+2\sum_r \breve{W}_{|} \Delta_i^{s(1)}\mu^{r(0)} \right),\cr
{\color{blue}\dot{\mu}^{s(1)}}=-\Lambda \Bigg( 2\breve{V}\mu^{s(1)}+2\sum_r \breve{W}_{|}\mu^{r(0)}\mu^{s(1)}+2\sum_r \breve{W}_{|}\mu^{r(1)}\mu^{s(0)}+2\sum_r \breve{W}_{|a}\Delta_a^{r(1)}\mu^{s(0)}+\cr
+ 2\breve{V}_a \Delta^{s(1)}_a+ 2\sum_r \breve{W}_{a|}\Delta^{s(1)}_a\mu^{r(0)}\Bigg).
\end{gathered}
\label{hohes5}
\end{equation}
\clearpage

The second order $K$-particle Hamilton--Ehrenfest system (accurate to $\Or(\hbar)$) includes \eqref{hohes4}, \eqref{hohes5}, and the following equations:
\begin{equation}
\begin{gathered}
{\color{blue}\dot{\Delta}^{s(2)}_{i}}=J_{ia}\sum_r W_{a|}\mu^{r(1)}\mu^{s(1)}+J_{ia}\sum_r W_{a|}\mu^{r(2)}\mu^{s(0)}+J_{ia}\sum_r W_{a|b}\Delta^{r(1)}_b\mu^{s(1)}+\\
+J_{ia}\sum_r W_{a|b}\Delta^{r(2)}_b\mu^{s(0)}+\dac{1}{2}J_{ia}\sum_r W_{a|bc}\Delta^{r(2)}_{bc}\mu^{s(0)}+\\
+J_{ia}V_{ab}\Delta_b^{s(2)}+J_{ia} \sum_r W_{ab|}\Delta_b^{s(2)}\mu^{r(0)}+J_{ia} \sum_r W_{ab|}\Delta_b^{s(1)}\mu^{r(1)}+J_{ia} \sum_r W_{ab|c}\Delta_b^{s(1)}\Delta_c^{r(1)}+\\
+\dac{1}{2}J_{ia}V_{abc}\Delta^{s(2)}_{bc}+\dac{1}{2}J_{ia}\sum_r W_{abc|}\Delta^{s(2)}_{bc} \mu^{r(0)}-\\
-\Lambda\Bigg( 2\breve{V}\Delta_i^{s(2)}+2\sum_r\breve{W}_{|} \Delta_i^{s(2)} \mu^{r(0)}+2\sum_r\breve{W}_{|} \Delta_i^{s(1)} \mu^{r(1)}+2\sum_r\breve{W}_{|a} \Delta_i^{s(1)} \Delta_a^{r(1)}+\\
+ 2\breve{V}_a \Delta_{ai}^{s(2)}+2\sum_r \breve{W}_{a|} \Delta_{ai}^{s(2)} \mu^{r(0)}\Bigg),\\
{\color{blue}\dot{\Delta}^{s(2)}_{ij}}=2J_{ia}\sum_r W_{a|}\Delta_j^{s(1)}\mu^{r(1)}+2J_{ia}\sum_r W_{a|b}\Delta_j^{s(1)}\Delta_b^{r(1)}+2J_{ia}V_{ab}\Delta_{bj}^{s(2)}+2J_{ia}\sum_r W_{ab|}\Delta_{bj}^{s(2)}\mu^{r(0)}-\\
-\Lambda\Bigg(2\breve{V} \Delta_{ij}^{s(2)}+2\sum_r \breve{W}_{|} \Delta_{ij}^{s(2)}\mu^{r(0)}\Bigg),\\
{\color{blue}\dot{\mu}^{s(2)}}=-\Lambda \Bigg( 2\breve{V}\mu^{s(2)}+2\sum_r\breve{W}_{|}\mu^{s(2)}\mu^{r(0)}+2\sum_r\breve{W}_{|}\mu^{s(1)}\mu^{r(1)}+2\sum_r\breve{W}_{|}\mu^{s(0)}\mu^{r(2)}+\\
+2\sum_r\breve{W}_{|a}\mu^{s(1)}\Delta_a^{r(1)}+2\sum_r\breve{W}_{|a}\mu^{s(0)}\Delta_a^{r(2)}+\sum_r\breve{W}_{|ab}\mu^{s(0)}\Delta_{ab}^{r(2)}+\\
+2\breve{V}_a \Delta_a^{s(2)}+2\sum_r\breve{W}_{a|}\Delta_a^{s(1)}\mu^{r(1)}+2\sum_r\breve{W}_{a|}\Delta_a^{s(2)}\mu^{r(0)}+2\sum_r\breve{W}_{a|b}\Delta_a^{s(1)}\Delta_b^{r(1)}+\\
+\breve{V}_{ab}\Delta_{ab}^{s(2)}+\sum_r \breve{W}_{ab|}\Delta_{ab}^{s(2)}\mu^{r(0)} \Bigg).
\end{gathered}
\label{hohes6}
\end{equation}

The third order $K$-particle Hamilton--Ehrenfest system (accurate to $\Or(\hbar^{3/2})$) includes \eqref{hohes4}, \eqref{hohes5}, \eqref{hohes6}, and the following equations:
\clearpage
\begin{equation}
\begin{gathered}
{\color{blue}\dot{\Delta}^{s(3)}_{i}}=J_{ia}\sum_r W_{a|}\mu^{r(1)}\mu^{s(2)}+J_{ia}\sum_r W_{a|}\mu^{r(2)}\mu^{s(1)}+J_{ia}\sum_r W_{a|}\mu^{r(3)}\mu^{s(0)}+\\
+J_{ia}\sum_r W_{a|b}\Delta_b^{r(1)}\mu^{s(2)}+J_{ia}\sum_r W_{a|b}\Delta_b^{r(2)}\mu^{s(1)}+J_{ia}\sum_r W_{a|b}\Delta_b^{r(3)}\mu^{s(0)}+\\
+\dac{1}{2}J_{ia}\sum_r W_{a|bc}\Delta_{bc}^{r(2)}\mu^{s(1)}+\dac{1}{2}J_{ia}\sum_r W_{a|bc}\Delta_{bc}^{r(3)}\mu^{s(0)}+\dac{1}{6}J_{ia}\sum_r W_{a|bcd}\Delta_{bcd}^{r(3)}\mu^{s(0)}+\\
+J_{ia}V_{ab}\Delta_b^{s(3)}+J_{ia}\sum_r W_{ab|}\mu^{r(0)}\Delta_b^{s(3)}+J_{ia}\sum_r W_{ab|}\mu^{r(1)}\Delta_b^{s(2)}+J_{ia}\sum_r W_{ab|}\mu^{r(2)}\Delta_b^{s(1)}+\\
+J_{ia}\sum_r W_{ab|c}\Delta_c^{r(1)}\Delta_b^{s(2)}+J_{ia}\sum_r W_{ab|c}\Delta_c^{r(2)}\Delta_b^{s(1)}+\dac{1}{2}J_{ia}\sum_r W_{ab|cd}\Delta_{cd}^{r(2)}\Delta_b^{s(1)}+\\
+\dac{1}{2}J_{ia}V_{abc}\Delta_{bc}^{s(3)}+\dac{1}{2}J_{ia}\sum_r W_{abc|}\mu^{r(1)}\Delta_{bc}^{s(2)}+\dac{1}{2}J_{ia}\sum_r W_{abc|}\mu^{r(0)}\Delta_{bc}^{s(3)}+\\
+\dac{1}{2}J_{ia}\sum_r W_{abc|d}\Delta_d^{r(1)}\Delta_{bc}^{s(2)}+\dac{1}{6}J_{ia}V_{abcd}\Delta_{bcd}^{s(3)}+\dac{1}{6}J_{ia}\sum_r W_{abcd|}\mu^{r(0)}\Delta_{bcd}^{s(3)}-\\
-\Lambda\Bigg(2\breve{V} \Delta_i^{s(3)}+2\sum_r \breve{W}_{|}\mu^{r(0)}\Delta_i^{s(3)}+2\sum_r \breve{W}_{|}\mu^{r(1)}\Delta_i^{s(2)}+2\sum_r \breve{W}_{|}\mu^{r(2)}\Delta_i^{s(1)}+\\
+2\sum_r \breve{W}_{|a}\Delta_a^{r(1)}\Delta_i^{s(2)}+2\sum_r \breve{W}_{|a}\Delta_a^{r(2)}\Delta_i^{s(1)}+\sum_r \breve{W}_{|ab}\Delta_{ab}^{r(2)}\Delta_i^{s(1)}+\\
+2\breve{V}_a \Delta_{ai}^{s(3)}+2\sum_r \breve{W}_{a|}\mu^{r(0)}\Delta_{ai}^{s(3)}+2\sum_r \breve{W}_{a|}\mu^{r(1)}\Delta_{ai}^{s(2)}+2\sum_r \breve{W}_{a|b}\Delta_b^{r(1)}\Delta_{ai}^{s(2)}+\\
+\breve{V}_{ab}\Delta_{abi}^{s(3)}+\sum_r \breve{W}_{ab|}\mu^{r(0)}\Delta_{abi}^{s(3)}\Bigg),\\
{\color{blue}\dot{\Delta}^{s(3)}_{ij}}=2J_{ia}\sum_r W_{a|}\mu^{r(1)}\Delta_j^{s(2)}+2J_{ia}\sum_r W_{a|}\mu^{r(2)}\Delta_j^{s(1)}+2J_{ia}\sum_r W_{a|b}\Delta_b^{r(1)}\Delta_j^{s(2)}+\\
+2J_{ia}\sum_r W_{a|b}\Delta_b^{r(2)}\Delta_j^{s(1)}+J_{ia}\sum_r W_{a|bc}\Delta_{bc}^{r(2)}\Delta_j^{s(1)}+\\
+2J_{ia}V_{ab}\Delta_{bj}^{s(3)}+2J_{ia}\sum_r W_{ab|}\mu^{r(0)}\Delta_{bj}^{s(3)}+2J_{ia}\sum_r W_{ab|}\mu^{r(1)}\Delta_{bj}^{s(2)}+\\
+2J_{ia}\sum_r W_{ab|c}\Delta_c^{r(1)}\Delta_{bj}^{s(2)}+J_{ia}V_{abc}\Delta_{bcj}^{s(3)}+J_{ia}\sum_r W_{abc|}\mu^{r(0)}\Delta_{bcj}^{s(3)}-\\
-\Lambda \Bigg( 2\breve{V}\Delta_{ij}^{s(3)}+2\sum_r \breve{W}_{|}\mu^{r(0)}\Delta_{ij}^{s(3)}+2\sum_r \breve{W}_{|}\mu^{r(1)}\Delta_{ij}^{s(2)}+\\
+2\sum_r \breve{W}_{|a}\Delta_a^{r(1)}\Delta_{ij}^{s(2)}+2\breve{V}_{a}\Delta_{aij}^{s(3)}+2\sum_r \breve{W}_{a|}\mu^{r(0)}\Delta_{aij}^{s(3)} \Bigg),\\
{\color{blue}\dot{\Delta}_{ijk}^{s(3)}}=3J_{ia}\sum_r W_{a|}\mu^{r(1)}\Delta_{jk}^{s(2)}+3J_{ia}\sum_r W_{a|b}\Delta_b^{r(1)}\Delta_{jk}^{s(2)}+\\
+3J_{ia}V_{ab}\Delta_{jkb}^{s(3)}+3J_{ia}\sum_r W_{ab|}\mu^{r(0)}\Delta_{jkb}^{s(3)}-\Lambda\Bigg(2\breve{V}\Delta_{ijk}^{s(3)}+2\sum_r \breve{W}_{|}\mu^{r(0)}\Delta_{ijk}^{s(3)}\Bigg).
\end{gathered}
\label{hohes6b}
\end{equation}

\begin{equation}
\begin{gathered}
{\color{blue}\dot{\mu}^{s(3)}}=\Lambda\Bigg(2\breve{V} \mu^{s(3)}+2\sum_r \breve{W}_{|}\mu^{r(0)}\mu^{s(3)}+2\sum_r \breve{W}_{|}\mu^{r(1)}\mu^{s(2)}+2\sum_r \breve{W}_{|}\mu^{r(2)}\mu^{s(1)}+\\
+2\sum_r \breve{W}_{|}\mu^{r(3)}\mu^{s(0)}+2\sum_r \breve{W}_{|a}\Delta_a^{r(1)}\mu^{s(2)}+2\sum_r \breve{W}_{|a}\Delta_a^{r(2)}\mu^{s(1)}+2\sum_r \breve{W}_{|a}\Delta_a^{r(3)}\mu^{s(0)}+\\
+\sum_r \breve{W}_{|ab}\Delta_{ab}^{r(2)}\mu^{s(1)}+\sum_r \breve{W}_{|ab}\Delta_{ab}^{r(3)}\mu^{s(0)}+\dac{1}{3}\sum_r \breve{W}_{|abc}\Delta_{abc}^{r(3)}\mu^{s(0)}+\\
+2\breve{V}_a\Delta_a^{s(3)}+2\sum_r \breve{W}_{a|}\mu^{r(0)}\Delta_a^{s(3)}+2\sum_r \breve{W}_{a|}\mu^{r(1)}\Delta_a^{s(2)}+2\sum_r \breve{W}_{a|}\mu^{r(2)}\Delta_a^{s(1)}+\\
+2\sum_r \breve{W}_{a|b}\Delta_b^{r(1)}\Delta_a^{s(2)}+2\sum_r \breve{W}_{a|b}\Delta_b^{r(2)}\Delta_a^{s(1)}+\sum_r \breve{W}_{a|bc}\Delta_{bc}^{r(2)}\Delta_a^{s(1)}+\\
+\breve{V}_{ab}\Delta_{ab}^{s(3)}+\sum_r \breve{W}_{ab|}\mu^{r(0)}\Delta_{ab}^{s(3)}+\sum_r \breve{W}_{ab|}\mu^{r(1)}\Delta_{ab}^{s(2)}+\sum_r \breve{W}_{ab|c}\Delta_c^{r(1)}\Delta_{ab}^{s(2)}+\\
+\dac{1}{3}\breve{V}_{abc}\Delta_{abc}^{s(3)}+\dac{1}{3}\sum_r \breve{W}_{abc|}\mu^{r(0)}\Delta_{abc}^{s(3)}\Bigg).
\end{gathered}
\label{hohes6a}
\end{equation}

From \eqref{hohes5}, \eqref{hohes6b},\eqref{hohes6a}, one readily gets that they admit solutions
\begin{equation}
\mu^{(2k+1)}_s(t,\BFC)= 0, \qquad \Delta_{s,i_1...i_m}^{(2k+1)}(t,\BFC)= 0, \qquad k\in {\mathbb{Z}_+},
\label{hohes7}
\end{equation}
for certain chose of initial conditions $\BFC$. Note that is also valid for the case when symbols $V$, $W$, $\breve{V}$, and $\breve{W}$ regularly depend on $\hbar$. In this work, we consider the case when $V$, $W$, $\breve{V}$, and $\breve{W}$ do not depend on the parameter $\hbar$ just for reasons of the brevity of formulae and simplicity of text. However, the generalization on the case of regular dependence of $V(z,t,\hbar)$, $W(z,w,t,\hbar)$, $\breve{V}(z,t,\hbar)$, and $\breve{W}(z,w,t,\hbar)$ on $\hbar$ does not meet the fundamental difficulties and just makes formulae \eqref{hohes6}---\eqref{hohes6a} more cumbersome.

\clearpage

If we consider the initial conditions for the equation \eqref{hartree1} that satisfy \eqref{hohes7} (for $t=0$), the system \eqref{hohes6} reads:
\begin{equation}
\begin{gathered}
{\color{blue}\dot{\Delta}^{s(2)}_{i}}=J_{ia}\sum_r W_{a|}\mu^{r(2)}\mu^{s(0)}+J_{ia}\sum_r W_{a|b}\Delta^{r(2)}_b\mu^{s(0)}+\dac{1}{2}J_{ia}\sum_r W_{a|bc}\Delta^{r(2)}_{bc}\mu^{s(0)}+\\
+J_{ia}V_{ab}\Delta_b^{s(2)}+J_{ia} \sum_r W_{ab|}\Delta_b^{s(2)}\mu^{r(0)}+\dac{1}{2}J_{ia}V_{abc}\Delta^{s(2)}_{bc}+\dac{1}{2}J_{ia}\sum_r W_{abc|}\Delta^{s(2)}_{bc} \mu^{r(0)}-\\
-\Lambda\Bigg( 2\breve{V}\Delta_i^{s(2)}+2\sum_r\breve{W}_{|} \Delta_i^{s(2)} \mu^{r(0)}+ 2\breve{V}_a \Delta_{ai}^{s(2)}+2\sum_r \breve{W}_{a|} \Delta_{ai}^{s(2)} \mu^{r(0)}\Bigg),\\
{\color{blue}\dot{\Delta}^{s(2)}_{ij}}=2J_{ia}V_{ab}\Delta_{bj}^{s(2)}+2J_{ia}\sum_r W_{ab|}\Delta_{bj}^{s(2)}\mu^{r(0)}-\Lambda\Bigg(2\breve{V} \Delta_{ij}^{s(2)}+2\sum_r \breve{W}_{|} \Delta_{ij}^{s(2)}\mu^{r(0)}\Bigg),\\
{\color{blue}\dot{\mu}^{s(2)}}=-\Lambda \Bigg( 2\breve{V}\mu^{s(2)}+2\sum_r\breve{W}_{|}\mu^{s(2)}\mu^{r(0)}+2\sum_r\breve{W}_{|}\mu^{s(0)}\mu^{r(2)}+\\
+2\sum_r\breve{W}_{|a}\mu^{s(0)}\Delta_a^{r(2)}+\sum_r\breve{W}_{|ab}\mu^{s(0)}\Delta_{ab}^{r(2)}+2\breve{V}_a \Delta_a^{s(2)}+\\
+2\sum_r\breve{W}_{a|}\Delta_a^{s(2)}\mu^{r(0)}+\breve{V}_{ab}\Delta_{ab}^{s(2)}+\sum_r \breve{W}_{ab|}\Delta_{ab}^{s(2)}\mu^{r(0)} \Bigg).
\end{gathered}
\label{hohes8}
\end{equation}
Also, $\mu^{s(1)}=\mu^{s(3)}=0$, $\Delta_i^{s(1)}=\Delta_i^{s(3)}=0$, $\Delta_{ij}^{(3)}=0$, $\Delta_{ijk}^{(3)}=0$ in this case. Therefore, if one limits himself to the initial conditions $\psi(\vec{x},\hbar)$ satisfying
\begin{equation}
\mu^{(2k+1)}(t)[\psi_s]\Big|_{t=0}= 0, \qquad \Delta_{i_1...i_m}^{(2k+1)}(t)[\psi_s]\Big|_{t=0}= 0, \qquad k=0,1, \qquad m=1,2,3.
\label{hohes7}
\end{equation}
the system \eqref{hohes4} is both the zeroth and first order Hamilton--Ehrenfest system (accurate to moments that are equal to zero) while the system \eqref{hohes8} on solutions to \eqref{hohes4} is both the second and third order Hamilton--Ehrenfest system. Note that the third order Hamilton--Ehrenfest system allows one to construct the first correction to the leading term of asymptotics, which has the same formal accuracy with respect to $\hbar$ as the leading term of asymptotics for the wave function constructed by the Maslov canonical operator method \cite{Maslov2} for the case $\varkappa=0$, $\Lambda=0$.

\bibliography{lit1}

\begin{thebibliography}{10}
\newcommand{\enquote}[1]{``#1''}
\expandafter\ifx\csname url\endcsname\relax
  \def\url#1{\texttt{#1}}\fi
\ifx \dourl  \undefined \def \dourl#1{\url{https://doi.org/#1}}\fi

\expandafter\ifx\csname urlprefix\endcsname\relax\def\urlprefix{URL }\fi
\providecommand{\eprint}[2][]{\url{#2}}

\bibitem{pitaevskii1999}
F.~Dalfovo, S.~Giorgini, L.~P. Pitaevskii, and S.~Stringari, \enquote{Theory of
  Bose-Einstein condensation in trapped gases,} Reviews of Modern Physics
  \textbf{71}(3), 463--512 (1999).

\bibitem{ashida20}
Y.~Ashida, Z.~Gong, and M.~Ueda, \enquote{Non-Hermitian physics,} Advances in
  Physics \textbf{69}, 249--435 (2020). \dourl{10.1080/00018732.2021.1876991}.

\bibitem{lederer2008}
F.~Lederer, G.~Stegeman, D.~Christodoulides, G.~Assanto, M.~Segev, and
  Y.~Silberberg, \enquote{Discrete solitons in optics,} Physics Reports-review
  Section of Physics Letters \textbf{463}, 1--126 (2008).
  \dourl{10.1016/j.physrep.2008.04.004}.

\bibitem{agrawal12}
G.~Agrawal, \emph{Nonlinear Fiber Optics} (2013).
  \dourl{10.1016/C2011-0-00045-5}.

\bibitem{aleksic21}
B.~Aleksi\'{c}, L.~Uvarova, and N.~Aleksi\'{c}, \enquote{Dissipative structures
  in the resonant interaction of laser radiation with nonlinear dispersive
  medium,} Optical and Quantum Electronics \textbf{53} (2021).
  \dourl{10.1007/s11082-021-03017-4}.

\bibitem{aleksic20}
B.~Aleksi\'{c}, L.~Uvarova, N.~Aleksi\'{c}, and M.~Beli\'{c}, \enquote{Cubic
  quintic Ginzburg Landau equation as a model for resonant interaction of EM
  field with nonlinear media,} Optical and Quantum Electronics \textbf{52}
  (2020). \dourl{10.1007/s11082-020-02271-2}.

\bibitem{fetter01}
A.~L. Fetter and A.~A. Svidzinsky, \enquote{Vortices in a trapped dilute
  Bose-Einstein condensate,} Journal of Physics Condensed Matter
  \textbf{13}(12), 135--194 (2001).

\bibitem{arecchi2000}
F.~Arecchi, J.~Bragard, and L.~Castellano, \enquote{Dissipative dynamics of an
  open Bose Einstein condensate,} Optics Communications \textbf{179}(1),
  149--156 (2000). \dourl{10.1016/S0030-4018(99)00670-7}.

\bibitem{haus1984}
H.~Haus, \emph{Waves and Fields in Optoelectronics}, Prentice-Hall series in
  solid state physical electronics (Prentice-Hall, 1984).

\bibitem{baranov2008}
M.~A. Baranov, \enquote{Theoretical progress in many-body physics with
  ultracold dipolar gases,} Physics Reports \textbf{464}(3), 71--111 (2008).

\bibitem{malomed2009}
J.~Cuevas, B.~A. Malomed, P.~G. Kevrekidis, and D.~J. Frantzeskakis,
  \enquote{Solitons in quasi-one-dimensional Bose-Einstein condensates with
  competing dipolar and local interactions,} Physical Review A - Atomic,
  Molecular, and Optical Physics \textbf{79}(5) (2009).

\bibitem{klaus2022}
L.~Klaus, T.~Bland, E.~Poli, C.~Politi, G.~Lamporesi, E.~Casotti, R.~Bisset,
  M.~Mark, and F.~Ferlaino, \enquote{Observation of vortices and vortex stripes
  in a dipolar condensate,} Nature Physics \textbf{18}, 1--6 (2022).
  \dourl{10.1038/s41567-022-01793-8}.

\bibitem{zhao2021}
Q.~Zhao, \enquote{Effects of Dipole-Dipole Interaction on Vortex Motion in
  Bose-Einstein Condensates,} Journal of Low Temperature Physics \textbf{204},
  1--11 (2021). \dourl{10.1007/s10909-021-02594-8}.

\bibitem{nizette21}
M.~Nizette and A.~Vladimirov, \enquote{Generalized Haus master equation model
  for mode-locked class- B lasers,} Physical Review E \textbf{104}, 014215
  (2021). \dourl{10.1103/PhysRevE.104.014215}.

\bibitem{curtis2012}
C.~W. Curtis, \enquote{On nonlocal Gross-Pitaevskii equations with periodic
  potentials,} Journal of Mathematical Physics \textbf{53}(7), 073709 (2012).

\bibitem{kout2024}
G.~Koutsokostas, I.~Moseley, T.~Horikis, and D.~Frantzeskakis,
  \enquote{Particle and wave dynamics of nonlocal solitons in external
  potentials,} Physics Letters A p. 129683 (2024).
  \dourl{10.1016/j.physleta.2024.129683}.

\bibitem{breev2022}
A.~Breev, A.~Shapovalov, and D.~Gitman, \enquote{Noncommutative Reduction of
  Nonlinear Schr\"{o}dinger Equation on Lie Groups,} Universe \textbf{8}, 445
  (2022). \dourl{10.3390/universe8090445}.

\bibitem{bobmann23}
L.~Bobmann, C.~Dietze, and P.~Nam, \enquote{Focusing dynamics of 2D Bose gases
  in the instability regime,}  (2023). \dourl{10.48550/arXiv.2307.00956}.

\bibitem{boccato15}
C.~Boccato, S.~Cenatiempo, and B.~Schlein, \enquote{Quantum Many-Body
  Fluctuations Around Nonlinear Schr\"{o}dinger Dynamics,} Annales Henri
  Poincar\`e \textbf{18}, 113--191 (2017). \dourl{10.1007/s00023-016-0513-6}.

\bibitem{benedikter14}
N.~Benedikter, G.~Oliveira, and B.~Schlein, \enquote{Quantitative Derivation of
  the Gross-Pitaevskii Equation,} Communications on Pure and Applied
  Mathematics \textbf{68}(8), 1399--1482 (2014). \dourl{10.1002/cpa.21542}.

\bibitem{pickl11}
P.~Pickl, \enquote{A Simple Derivation of Mean Field Limits for Quantum
  Systems,} Letters in Mathematical Physics \textbf{97}, 151--164 (2011).
  \dourl{10.1007/s11005-011-0470-4}.

\bibitem{spohn80}
H.~Spohn, \enquote{Kinetic equations from Hamiltonian dynamics: Markovian
  limits,} Rev. Mod. Phys. \textbf{52}, 569--616 (1980).
  \dourl{10.1103/RevModPhys.52.569}.

\bibitem{marcucci19}
G.~Marcucci, D.~Pierangeli, S.~Gentilini, N.~Ghofraniha, Z.~Chen, and C.~Conti,
  \enquote{Optical spatial shock waves in nonlocal nonlinear media,} Advances
  in Physics: X \textbf{4}, 1662733 (2019).
  \dourl{10.1080/23746149.2019.1662733}.

\bibitem{santos16}
F.~W\"achtler and L.~Santos, \enquote{Quantum filaments in dipolar
  Bose-Einstein condensates,} Phys. Rev. A \textbf{93}, 061603 (2016).
  \dourl{10.1103/PhysRevA.93.061603}.

\bibitem{Maslov2}
V.~Maslov, \emph{The Complex WKB Method for Nonlinear Equations. I. Linear
  Theory} (Birkhauser Verlag, Basel, 1994).

\bibitem{BeD2}
V.~V. Belov and S.~Y. Dobrokhotov, \enquote{Semiclassical Maslov asymptotics
  with complex phases. I. General approach,} Theoretical and Mathematical
  Physics \textbf{92}(2), 843--868 (1992).

\bibitem{shapovalov:BTS1}
V.~V. Belov, A.~Y. Trifonov, and A.~V. Shapovalov, \enquote{The
  trajectory-coherent approximation and the system of moments for the hartree
  type equation,} International Journal of Mathematics and Mathematical
  Sciences \textbf{32}(6), 325--370 (2002).

\bibitem{lisok2007}
A.~L. Lisok, A.~Y. Trifonov, and A.~V. Shapovalov, \enquote{Quasi-energy
  spectral series for a nonlocal Gross-Pitaevskii equation,} Russian Physics
  Journal \textbf{50}(7), 695--709 (2007).

\bibitem{athanas11}
A.~Athanassoulis, T.~Paul, F.~Pezzotti, and M.~Pulvirenti,
  \enquote{Semiclassical Propagation of Coherent States for the Hartree
  Equation,} Ann. Henri Poincare \textbf{22}, 1613--1634 (2011).
  \dourl{10.1007/s00023-011-0115-2}.

\bibitem{Pereskokov2017}
A.~V. Pereskokov, \enquote{Semiclassical asymptotics of the spectrum near the
  lower boundary of spectral clusters for a Hartree-type operator,}
  Mathematical Notes \textbf{101}(5-6), 1009--1022 (2017).

\bibitem{pereskokov24}
A.~Pereskokov, \enquote{Asymptotics of the Spectrum of a Hartree Type Operator
  with Self-Consistent Potential Including the Macdonald Function,} Journal of
  Mathematical Sciences \textbf{279}, 1--17 (2024).
  \dourl{10.1007/s10958-024-07029-9}.

\bibitem{kulagin2024}
A.~E. Kulagin and A.~V. Shapovalov, \enquote{A Semiclassical Approach to the
  Nonlocal Nonlinear Schr\"{o}dinger Equation with a Non-Hermitian Term,}
  Mathematics \textbf{12}(4), 580 (2024). \dourl{10.3390/math12040580}.

\bibitem{shi2023}
Z.~Shi, F.~Badshah, L.~Qin, Y.~Zhou, H.~Huang, and Y.~Zhang, \enquote{Spatially
  modulated control of pattern formation in a general nonlocal nonlinear
  system,} Chaos, Solitons and Fractals \textbf{175}, 113929 (2023).
  \dourl{10.1016/j.chaos.2023.113929}.

\bibitem{saito16}
K.-T. Xi and H.~Saito, \enquote{Droplet formation in a Bose-Einstein condensate
  with strong dipole-dipole interaction,} Physical Review A \textbf{93}, 011604
  (2016). \dourl{10.1103/PhysRevA.93.011604}.

\bibitem{kulshap24}
A.~E. Kulagin and A.~V. Shapovalov, \enquote{Quasiparticles for the
  one-dimensional nonlocal Fisher-Kolmogorov-Petrovskii-Piskunov equation,}
  Physica Scripta \textbf{99}(4), 045228 (2024).
  \dourl{10.1088/1402-4896/ad302c}.

\bibitem{zakharov84}
V.~E. Zakhavrov, S.~V. Manakov, S.~P. Novikov, and L.~P. Pitaevskii,
  \emph{Theory of solitons. The inverse scattering method} (Consultants Bureau
  Platform, NY, 1984).

\bibitem{dodd82}
R.~Dodd, J.~Eilbeck, J.~Gibbon, and H.~Morris, \emph{Soliton and nonlinear wave
  equations} (Academic Press, London, 1982).

\bibitem{karpman81}
V.~Karpman and V.~Solov'ev, \enquote{A perturbation theory for soliton
  systems,} Physica D: Nonlinear Phenomena \textbf{3}(1), 142--164 (1981).
  \dourl{10.1016/0167-2789(81)90123-8}.

\bibitem{ribeiro23}
C.~Ribeiro and U.~Fischer, \enquote{Nonlocal field theory of quasiparticle
  scattering in dipolar Bose-Einstein condensates,} SciPost Physics Core
  \textbf{6} (2023). \dourl{10.21468/SciPostPhysCore.6.1.003}.

\bibitem{Maslov1}
V.~Maslov, \emph{Operational Methods} (Mir Publishers, Moscow, 1976).

\bibitem{maslov81}
V.~Maslov and V.~Nazaikinskii, \enquote{Algebras with general commutation
  relations and their applications. I. Pseudodifferential equations with
  increasing coefficients,} Journal of Mathematical Sciences \textbf{15},
  167--273 (1981). \dourl{10.1007/BF01083678}.

\bibitem{sym2020}
A.~V. Shapovalov, A.~E. Kulagin, and A.~Y. Trifonov, \enquote{The
  Gross--Pitaevskii equation with a nonlocal interaction in a semiclassical
  approximation on a curve,} Symmetry \textbf{12}(2), 201 (2020).
  \dourl{10.3390/sym12020201}.

\bibitem{choi98}
S.~Choi, S.~A. Morgan, and K.~Burnett, \enquote{Phenomenological damping in
  trapped atomic Bose-Einstein condensates,} Phys. Rev. A \textbf{57},
  4057--4060 (1998). \dourl{10.1103/PhysRevA.57.4057}.

\bibitem{obukhov2022}
V.~Obukhov, \enquote{Algebras of integrals of motion for the Hamilton-Jacobi
  and Klein-Gordon-Fock equations in spacetime with four-parameter groups of
  motions in the presence of an external electromagnetic field,} Journal of
  Mathematical Physics \textbf{63}, 023505 (2022). \dourl{10.1063/5.0080703}.

\bibitem{osetrin2021}
K.~Osetrin, I.~Kirnos, E.~Osetrin, and A.~Filippov, \enquote{Wave-Like Exact
  Models with Symmetry of Spatial Homogeneity in the Quadratic Theory of
  Gravity with a Scalar Field,} Symmetry \textbf{13}, 1173 (2021).
  \dourl{10.3390/sym13071173}.

\bibitem{multiind}
M.~V. Karasev, \enquote{Weyl and ordered calculus of noncommuting operators,}
  Mathematical Notes of the Academy of Sciences of the USSR \textbf{26}(6),
  945--958 (1979).

\end{thebibliography}

\end{document}